\pdfoutput=1
\documentclass[aps,prl,10pt,twocolumn,superscriptaddress,nofootinbib,floatfix,longbibliography]{revtex4-2}
\usepackage[T1]{fontenc}
\usepackage{amsmath,amssymb,mathtools,amsthm}
\usepackage{mathrsfs}
\usepackage{graphicx,booktabs,xcolor,needspace}
\usepackage[colorlinks=true,allcolors=blue!60!black]{hyperref}
\DeclareMathOperator{\tr}{tr}
\DeclareMathOperator{\supp}{supp}

\newcommand{\Lin}{\mathsf{L}}
\newcommand{\SOC}{\mathsf{SOC}}
\newcommand{\T}{\mathsf{T}}
\newcommand{\Gsf}{\mathsf{G}}
\newcommand{\opt}{\mathrm{opt}}
\newcommand{\MLD}{\mathrm{MLD}}

\newcommand{\CCal}{\mathcal{C}}
\newcommand{\DCal}{\mathcal{D}}
\newcommand{\ECal}{\mathcal{E}}
\newcommand{\FCal}{\mathcal{F}}
\newcommand{\HCal}{\mathcal{H}}
\newcommand{\NCal}{\mathcal{N}}
\newcommand{\PCal}{\mathcal{P}}
\newcommand{\RCal}{\mathcal{R}}
\newcommand{\SCal}{\mathcal{S}}
\newcommand{\TCal}{\mathcal{T}}
\newcommand{\ftilde}{\widetilde{f}}
\newcommand{\ket}[1]{\lvert #1\rangle}
\newcommand{\bra}[1]{\langle #1\rvert}
\newcommand{\kb}[2]{\lvert #1\rangle\!\langle #2\rvert}
\newcommand{\dket}[1]{\lvert #1\rangle\!\rangle}
\newcommand{\norm}[1]{\lVert #1\rVert}
\newcommand{\proj}{\widehat{\Pi}}
\newtheorem{theorem}{Theorem}
\providecommand{\tr}{\operatorname{tr}}
\providecommand{\supp}{\operatorname{supp}}
\providecommand{\rank}{\operatorname{rank}}

\providecommand{\Lin}{\mathsf{L}}

\providecommand{\SOC}{\mathsf{SOC}}
\providecommand{\Gsf}{\mathsf{G}}
\providecommand{\opt}{\mathrm{opt}}
\providecommand{\Fe}{F_\mathrm{e}}
\providecommand{\MLD}{\mathrm{MLD}}
\providecommand{\ket}[1]{\lvert #1\rangle}
\providecommand{\bra}[1]{\langle #1\rvert}
\providecommand{\kb}[2]{\lvert #1\rangle\!\langle #2\rvert}
\providecommand{\dket}[1]{\lvert #1\rangle\!\rangle}
\providecommand{\dbra}[1]{\langle\!\langle #1\rvert}
\providecommand{\dbraket}[2]{\langle\!\langle #1\vert #2\rangle\!\rangle}
\providecommand{\norm}[1]{\lVert #1\rVert}
\newtheorem{smtheorem}{Theorem}

\newtheorem{smlemma}[smtheorem]{Lemma}

\theoremstyle{definition}

\theoremstyle{remark}

\theoremstyle{plain}
\makeatletter
\let\smtoc@section\l@section
\let\smtoc@subsection\l@subsection
\let\smtoc@subsubsection\l@subsubsection
\newcommand{\smtocstart}{%
  \let\l@section\smtoc@section
  \let\l@subsection\smtoc@subsection
  \let\l@subsubsection\smtoc@subsubsection
}
\newcommand{\supplementtableofcontents}{%
  \begingroup
  \let\l@section\@gobble@tw@
  \let\l@subsection\@gobble@tw@
  \let\l@subsubsection\@gobble@tw@
  \tableofcontents
  \endgroup
}
\makeatother
\begin{document}
\title{Advancing the Theory of Pretty Good Quantum Operations}
\author{Bikun Li}
\affiliation{Chicago Quantum Institute and Pritzker School of Molecular Engineering, University of Chicago, Chicago, Illinois 60637, USA}
\author{Guo Zheng}
\affiliation{Chicago Quantum Institute and Pritzker School of Molecular Engineering, University of Chicago, Chicago, Illinois 60637, USA}
\author{Liang Jiang}
\affiliation{Chicago Quantum Institute and Pritzker School of Molecular Engineering, University of Chicago, Chicago, Illinois 60637, USA}
\begin{abstract}
Finding optimal quantum measurements or recovery maps is often
demanding. Pretty good measurements and Petz maps offer explicit
approximations achieving at least the square of optimal performance by the
Barnum--Knill bound. We unify these operations in a semidefinite
programming framework with a tunable power, extending them to broader
quantum tasks. We prove a stronger bound for powers above one up to
three that peaks at quadratic power. For structured channels such as
stabilizer codes under Pauli noise, large powers steer recovery to a
maximum likelihood decoder for maximally mixed inputs.
\end{abstract}
\maketitle

\emph{Introduction.---}Quantum information theory often replaces a
difficult optimization with an explicit operation that is simple enough
to analyze and implement. The pretty good measurement~(PGM) is the
canonical example in state discrimination. Its square root normalization
balances prior probabilities against state overlaps without solving for
the optimal measurement~\cite{HausladenWootters1994,EldarForney2001}.
The Petz map, sometimes known as the pretty good recovery
map~\cite{BertaColesWehner2014,ItenRenesSutter2017}, plays the
corresponding role in quantum information recovery, notably in
approximate quantum error correction~\cite{BK2002,NgMandayam2010}.
It originally arose from equality in the data processing
inequality~\cite{Petz1986,Petz1988} and has many useful applications in
quantum information science~\cite{Sutter_2018,Kwon2022PRL,PhysRevA.96.022118,Penington2022}.

While many useful properties justify these
constructions~\cite{HausladenWootters1994,surace2026inevitabilitypetztransposemap},
the most straightforward common explanation for the name ``pretty good''
may come from the performance guarantee provided by the Barnum--Knill~(BK)
bound~\cite{BK2002}. For typical quantum state discrimination or recovery
tasks, the PGM or Petz map achieves at least the square of the optimal
success probability or entanglement fidelity. Near perfect performance,
the resulting error is at most approximately twice the optimal error.
Such approximately optimal performance has attracted increasing attention
in recent studies beyond traditional settings, including quantum ensembles
and shadow tomography~\cite{MishraLamiWilde2025,Grewal2026}.

Therefore, it is natural to ask two questions.
First, can one provide an improved ``pretty good'' construction that
retains this simplicity while obtaining a provably stronger performance
guarantee? Second, can one establish a general ``pretty good'' framework that applies to a broader class of quantum optimization problems?

A natural clue may be provided by spectral reweighting. Its origins
include Belavkin's weighted square-root measurements~\cite{Belavkin1975},
their least-squares formulation~\cite{EldarForney2001}, and Mochon's
family of generalized PGMs~\cite{Mochon2006}. Specific examples include
Holevo's quadratic weighting near orthogonality~\cite{Holevo1979} and cubic
power reweighting with postmeasurement information~\cite{BallesterWehnerWinter2008}.
Subsequent work studied power reweighting and established that
quadratic weighting satisfies the BK bound in discrimination, recovery,
and a general maximum overlap problem~\cite{TysonWeighted2009,Tyson2009,Tyson2010,Audenaert_2014}.
More recently, Zhou et al.~\cite{ZhouChessaChitambarLeditzky2025} showed
that the power PGM approaches an optimal measurement at infinite power
for certain geometrically uniform ensembles. These results suggest that the
square root construction belongs to a larger family, motivating a common
formulation and performance analysis across measurement, recovery, and
more general quantum optimization tasks.

We answer both questions using a common framework for semidefinite
programs~(SDPs) that covers a wide range of optimization problems in
quantum information.
Power reweighting unifies the power PGM, a power-reweighted Petz map~(PPM)
encompassing the ordinary Petz and quadratic recovery maps, prover response
maps, and resource witnesses. Our \textit{lifted BK bound} holds for all
$q\ge1$, strictly improves on the BK bound for $1<q\le3$, and peaks at
$q=2$, where a normalized optimum $1-\eta$ ($\eta\ll1$) has its leading
guaranteed error reduced from $2\eta$ to $5\eta/4$.

This work also extends the discussion of the power reweighting mechanism by
considering the \textit{sectored orthogonal channel}~(SOC) in quantum error
correction~(QEC). The SOC class covers many familiar settings in the QEC
community, such as stabilizer codes under Pauli noise.
For a maximally mixed input, the power-reweighted decoder converges to an optimal maximum likelihood
decoder~(MLD) as $q\to\infty$. The Supplemental Material~(SM) proves
the main bounds and gives counterexamples that delimit the universal
claims~\cite{SM}.

\begin{table*}[t]
\caption{Applications of the power-reweighted pretty good construction. The cited
equations specify each measurement, map, or witness.}
\label{tab:unification}
\renewcommand{\arraystretch}{1.20}
\setlength{\tabcolsep}{4pt}
\begin{tabular*}{\textwidth}{@{\extracolsep{\fill}}lll@{}}
\toprule
\raisebox{0.5\normalbaselineskip}{Task} & \shortstack[l]{Cost matrix $C$ and\\reference matrix $\zeta$}
 & \shortstack[l]{Power-reweighted construction\\($q\ge1$)}\\
\midrule
Optimizing state discrimination
 & $C=\bigoplus_i p_i\rho_i$, $\zeta_i=1$
 & Power PGM, Eq.~\eqref{eq:genpgm}\\
\addlinespace[0.6ex]
Optimizing the recovery map & $C=J_{\rho^2}(\ECal)$, $\zeta=\rho^{\T}$
 & PPM, Eq.~\eqref{eq:ppm}\\
\addlinespace[0.6ex]
Maximizing acceptance in a quantum interactive proof
 & $C=C_{\mathrm{ver}}$, $\zeta=\sqrt{\tr_B C}/\tr\sqrt{\tr_B C}$
 & Processing map $\SCal^{[q]}$, Eq.~\eqref{eq:verifier-prover}\\
\addlinespace[0.6ex]
Asymmetry quantum resource quantification
 & $C=V_{\Gsf}\rho V_{\Gsf}^\dagger$, $\zeta=\tau$
 & $W^{[q]}$, Eq.~\eqref{eq:asymmetry-witness}\\
\bottomrule
\end{tabular*}
\end{table*}

\emph{Pretty good constructions.---}Write $\Lin(\HCal_A)$ for the set of square linear operators on
$\HCal_A$. All Hilbert spaces are finite dimensional, with
$d_A\coloneqq\dim\HCal_A$. A quantum channel is a
completely positive, trace preserving linear map
$\ECal:\Lin(\HCal_A)\to\Lin(\HCal_B)$. The symbols $\dagger$ and
$\T$ denote the adjoint and transpose, respectively.
The adjoint map $\ECal^\dagger$ is defined by
$\tr[Y^\dagger\ECal(X)]=\tr[(\ECal^\dagger(Y))^\dagger X]$.
The unnormalized Choi matrix is
$J(\ECal)\coloneqq\sum_{a,b}\kb{a}{b}\otimes\ECal(\kb{a}{b})$~\cite{Choi1975},
with the input factor listed first. We write $\proj_\alpha$ for an
orthogonal projector and $\proj_M$ for the projector onto $\supp(M)$.
Real powers of a positive matrix act on its support, so
$M^0=\proj_M$ and $M^{-1}$ is its Moore--Penrose inverse.
For a positive input operator $\rho$, define the dressed Choi matrix
\begin{equation}\label{eq:dressed}
 J_\rho(\ECal)\coloneqq(\sqrt\rho^{\T}\otimes I_B)
 J(\ECal)(\sqrt\rho^{\T}\otimes I_B),
\end{equation}
to represent a purification of $\rho$ transformed locally by $\ECal$.

For a nonzero reference $\sigma\succeq0$ and a completely
positive~(CP) map $\FCal$, the Petz map is defined as
\begin{equation}\label{eq:petz}
 \PCal_{\sigma,\FCal}(\omega)\coloneqq\sqrt\sigma\FCal^\dagger
 \left[\FCal(\sigma)^{-1/2}\omega\FCal(\sigma)^{-1/2}\right]\sqrt\sigma.
\end{equation}
As a map from $\Lin(\HCal_B)$ to $\Lin(\HCal_A)$, it is trace
nonincreasing and is trace preserving on $\supp(\FCal(\sigma))$.
Throughout this work, $\tau\coloneqq I_A/d_A$ denotes the maximally mixed
input state.

For pure-state ensembles, a generalized PGM allows a separate weight
for each state~\cite{Belavkin1975,Mochon2006}. The power PGM uses a
common exponent $q$ for all ensemble operators~\cite{Audenaert_2014,ZhouChessaChitambarLeditzky2025}.
For an ensemble $\{(p_i,\rho_i)\}_i$ with $p_i>0$ and real $q$, set
$S_q\coloneqq\sum_j(p_j\rho_j)^q$. The resulting positive operator
valued measure~(POVM) has effects
\begin{equation}\label{eq:genpgm}
 M_i^{[q]}\coloneqq S_q^{-1/2}(p_i\rho_i)^q S_q^{-1/2}.
\end{equation}
When $q=1$, $S_1=\overline\rho\coloneqq\sum_i p_i\rho_i$, and these
effects give the ordinary PGM on $\supp(\overline\rho)$. Its measurement
map is $\PCal_{\sigma_{\mathrm{c}},\ECal_{\mathrm{cq}}}$, with prior state
$\sigma_{\mathrm{c}}\coloneqq\sum_i p_i\kb{i}{i}$ and preparation channel
$\ECal_{\mathrm{cq}}$ having Choi matrix
$J(\ECal_{\mathrm{cq}})\coloneqq\sum_i\kb{i}{i}\otimes\rho_i$. In the
classical Bayesian model $\rho_i=\sum_j p(j|i)\kb{j}{j}$, the power PGM
maximizes $p_{\mathrm{succ}}\coloneqq\sum_i p_i\tr(\rho_i M_i)$ as
$q\to\infty$, whereas general ensembles typically require solving an SDP.

\emph{The optimization problem family.---}To go beyond $p_{\mathrm{succ}}$, we now introduce the family of
SDPs relevant to the BK bound in this work. For a Hermitian cost matrix
$C\succeq0$ on $\HCal_{A}\otimes\HCal_{B}$, consider
\begin{equation}\label{eq:sdp}
 f^\opt\coloneqq\max\{\tr(CX):X\succeq0,\tr_A X\preceq I_B\}.
\end{equation}
For $C\succeq0$, replacing the inequality constraint by equality leaves the
optimum unchanged~\cite{SM}. Specializing $C$ adapts this SDP to concrete
tasks. In particular, to optimize a CP, trace nonincreasing recovery map
$\RCal:\Lin(\HCal_B)\to\Lin(\HCal_A)$ for the entanglement fidelity
$\Fe(\rho,\RCal\circ\ECal)$ at input state
$\rho$~\cite{Schumacher1996,Fletcher2007}, we take $A$ as the logical
system, $B$ as the output of the quantum channel $\ECal$, and
$X=J(\RCal^\dagger)$ as the adjoint Choi matrix of $\RCal$. Setting
$C=J_{\rho^2}(\ECal)$ then gives $\tr(CX)=\Fe(\rho,\RCal\circ\ECal)$.
For minimum-error state discrimination, the cost matrix is
$C=J_{\sigma_{\mathrm{c}}}(\ECal_{\mathrm{cq}})=\bigoplus_i p_i\rho_i$.
With the POVM effects $M_i$ as the diagonal blocks of $X$, the objective
$\tr(CX)$ equals the success probability $p_{\mathrm{succ}}$. Note that
when the objective $f\coloneqq\tr(CX)$ represents $p_{\mathrm{succ}}$ or
$\Fe$, the ordinary PGM and Petz map satisfy the well known BK bound
$f\ge(f^\opt)^2$.

This SDP also covers state transformation and average gate
fidelity~\cite{Fiurasek2001,Nielsen2002}, with costs and normalizations
for Table~\ref{tab:unification} derived in the SM~\cite{SM}. Depending on
$C$, $X$ represents a quantum map, resource witness, et al.
Generally, the solution of this SDP is purely numerical and lacks a closed form,
although the corresponding PGM and Petz map are known to be optimal under certain 
conditions~\cite{EldarMegretskiVerghese2004,DallaPozzaPierobon2015,ItenRenesSutter2017,Li2025}.

We highlight that Eq.~\eqref{eq:sdp} always admits an optimal solution
representable as a special Petz map. Let
$\HCal_{B,+}\coloneqq\supp(\tr_A C)\subseteq\HCal_B$ be the populated
output subspace. After restricting the output Hilbert space to
$\HCal_{B,+}$, we have $\tr_A C\succ0$, and global convergence of the
Reimpell--Werner iteration~(RW)~\cite{RW2005,Ru2026} under suitable
support conditions guarantees an optimizer $X^\opt$ that is a fixed point
of the RW iteration. Define the auxiliary CP map $\FCal^\opt$ by
$J(\FCal^\opt)\coloneqq CX^\opt C$. The fixed point identity then gives
$X^\opt=J((\PCal_{\tau,\FCal^\opt})^\dagger)$~\cite{SM}. However,
$\FCal^\opt$ depends on the unknown optimizer, which turns the general approximation into
the constructive task of choosing an explicit auxiliary map. Related uses
of auxiliary maps appear in Ref.~\cite{BiswasMandayam2025}.

\emph{Power reweighting.---}We now show how to tune the Petz map with a $q$-power auxiliary map, as a generalized from Eq.~\eqref{eq:genpgm}. Write the cost matrix in
the block diagonal form $C=\bigoplus_a C_a$, possibly with a single
block, where $C_a$ acts on
$\HCal_{A_a}\otimes\HCal_B$ and $\HCal_A=\bigoplus_a\HCal_{A_a}$.
For each nonzero cost block
$C_a$, choose a
nonzero reference matrix $\zeta_a\succeq0$ satisfying the admissibility conditions
$\tr \zeta_a\le1$ and $\supp(C_a)\subseteq\supp(\zeta_a)\otimes\HCal_{B}$.
\Needspace{8\baselineskip}
Let
\begin{equation}\label{eq:reference}
 \begin{gathered}
  Z\coloneqq\bigoplus_a(\zeta_a\otimes I_B),\quad H\coloneqq Z^{-1/2}CZ^{-1/2}\\
   c_\zeta\coloneqq\tr H=\tr(CZ^{-1}).
 \end{gathered}
\end{equation}
We call $c_\zeta$ the normalization cap because $f^\opt\le c_\zeta$, and
for a single cost block it is minimized by
$\zeta=\sqrt{\tr_B C}/\tr\sqrt{\tr_B C}$~\cite{SM}.

For each real power $q$, we define a positive seed $G_q$, its
normalizing factor $Y_q$, the feasible matrix $X_q$, and its objective
value $f^{[q]}$ by
\begin{subequations}\label{eq:power}
\begin{align}
 G_q&\coloneqq Z^{(q-1)/2}H^qZ^{(q-1)/2},\label{eq:power-seed}\\
 Y_q&\coloneqq\sqrt{\tr_A G_q},\label{eq:power-marginal}\\
 X_q&\coloneqq(I_A\otimes Y_q^{-1})G_q
 (I_A\otimes Y_q^{-1}),\label{eq:power-decoder}\\
 f^{[q]}&\coloneqq\tr(CX_q).\label{eq:power-objective}
\end{align}
\end{subequations}
Defining the auxiliary CP map $\widehat\FCal^{[q]}$ by
$J(\widehat\FCal^{[q]})\coloneqq G_q$, the factors $Y_q^{-1}$ enforce
the constraint in Eq.~\eqref{eq:sdp}, while $q$ controls how strongly $G_q$ amplifies
larger eigenvalues of $H$.

The associated power-reweighted construction $X_q$ has the Petz map
representation
\begin{equation}\label{eq:petz-representation}
 X_q=J\big(\big(\PCal_{\tau,\widehat\FCal^{[q]}}\big)^\dagger\big).
\end{equation}
Applying this formalism to entanglement fidelity gives a power-reweighted
version of the Petz map. Set $C=J_{\rho^2}(\ECal)$ and
$\zeta=\rho^{\T}$, so that $H=J_\rho(\ECal)$.
Using Tyson's state dependent functional calculus~\cite{Tyson2010},
define the CP map $\ECal^{(q,\rho)}$ by
$J(\ECal^{(q,\rho)})\coloneqq Z^{-1/2}H^qZ^{-1/2}$.
Then $Y_q^2=\ECal^{(q,\rho)}(\rho^q)$, and the associated PPM acts as
\begin{equation}\label{eq:ppm}
 \begin{gathered}
 \PCal_{\rho^q,\ECal^{(q,\rho)}}(\omega)\coloneqq\rho^{q/2}(\ECal^{(q,\rho)})^\dagger
 \bigl(Y_q^{-1}\omega Y_q^{-1}\bigr)\rho^{q/2}.
 \end{gathered}
\end{equation}
At $q=1$ this is the ordinary Petz map, and at $q=2$ it gives the
quadratic recovery map~\cite{Tyson2010}. 
At every power, the entanglement fidelity is evaluated at the original input state $\rho$.
We note that no direct conversion is known in general between the
PPM and rotated Petz maps, which use modular rotations~\cite{Wilde2017}.

For state discrimination, the cost blocks $C_i=p_i\rho_i$ and scalar
references $\zeta_i=1$ give seed blocks $(p_i\rho_i)^q$. Normalizing these
blocks recovers the effects of the power PGM in Eq.~\eqref{eq:genpgm}.
By contrast, for $q\ne1$, substituting $\rho=\sigma_{\mathrm{c}}$ and
$\ECal=\ECal_{\mathrm{cq}}$ directly into the PPM of Eq.~\eqref{eq:ppm}
generally yields a different POVM.

Two further examples illustrate this framework beyond discrimination and
recovery (details in the SM~\cite{SM}).
The first example is a two-message quantum interactive
proof~\cite{JainUpadhyayWatrous2009}. A verifier prepares a canonical
purification of a question state $\xi_B$, retains the purifying system (a
copy of $\HCal_B$), and sends the question system $B$ to a prover, who
returns an answer system $A$ through a quantum channel
$\SCal:\Lin(\HCal_B)\to\Lin(\HCal_A)$. The verifier then performs a joint
measurement on $A$ and the purifying system, accepting with POVM effect
$M_{\mathrm{acc}}$ on $\HCal_A\otimes\HCal_B$. For $X=J(\SCal^\dagger)$,
the acceptance probability is $\tr(C_{\mathrm{ver}}X)$ for the cost matrix
$C_{\mathrm{ver}}$ below, so $f^\opt$ in Eq.~\eqref{eq:sdp} is the maximum
acceptance probability~\cite{JainUpadhyayWatrous2009}. The cost matrix and power-reweighted processing map are
\begin{subequations}\label{eq:verifier-family}
\begin{align}
 C_{\mathrm{ver}}&\coloneqq(I_A\otimes\sqrt{\xi_B})M_{\mathrm{acc}}^{\T}
 (I_A\otimes\sqrt{\xi_B}),
 \label{eq:verifier-cost}\\
 \SCal^{[q]}&\coloneqq\PCal_{\tau,\widehat\FCal^{[q]}}.
 \label{eq:verifier-prover}
\end{align}
\end{subequations}
This map is trace nonincreasing and achieves acceptance probability
$f^{[q]}$. Additional Kraus operators acting on the orthogonal complement
of $\HCal_{B,+}$ can be appended to $\SCal^{[q]}$ to make it trace
preserving without changing the objective.

The second example comes from the resource theory of asymmetry, which also
illustrates that the optimization problems covered by Eq.~\eqref{eq:sdp}
do not involve physical bipartite systems $A$ and $B$ originally.
Consider any unitary representation $g\mapsto U_g$ of a finite or compact
group $\Gsf$ acting on a physical system $\HCal_B$. The group
twirling map on $\HCal_B$ is
$\TCal_{\Gsf}(\cdot)\coloneqq\int_{\Gsf}U_g(\cdot)U_g^\dagger\mathrm{d}\mu(g)$
with normalized Haar measure $\mu$~\cite{BartlettRudolphSpekkens2007}, and
the invariant states form the set
$\mathbb{F}_{\Gsf}=\{\sigma\succeq0:\tr\sigma=1,\TCal_{\Gsf}(\sigma)=\sigma\}$.
For a state $\rho$ on $\HCal_B$, the generalized robustness of asymmetry
is given by~\cite{Piani2016Asymmetry,TakagiRegulaWilde2022}
\begin{equation}\label{eq:asymmetry-definition}
 \begin{aligned}
 \mathscr{A}_{\mathrm{R}}(\rho)
 &=\min_{s\ge0,\sigma\in\mathbb{F}_{\Gsf}}
 \{s:\rho\preceq(1+s)\sigma\}\\
 &=\max_{W\succeq0}\{\tr(\rho W):\TCal_{\Gsf}(W)\preceq I_B\}-1.
 \end{aligned}
\end{equation}
Here $\HCal_A$ enters as an auxiliary environment holding the random
information of the twirling unitary, realizing $\TCal_{\Gsf}$ through
a Stinespring dilation with isometry $V_{\Gsf}:\HCal_B\to\HCal_A\otimes\HCal_B$
satisfying $\TCal_{\Gsf}(\cdot)=\tr_A[V_{\Gsf}(\cdot)V_{\Gsf}^\dagger]$.
For a finite group, one may take
$V_{\Gsf}=|\Gsf|^{-1/2}\sum_{g\in\Gsf}\ket{g}_A\otimes U_g$,
whereas a minimal Stinespring dilation applies to compact groups as well
and has environment dimension $d_A=\rank[J(\TCal_{\Gsf})]$, the Choi
matrix rank of $\TCal_{\Gsf}$. Setting
$C=V_{\Gsf}\rho V_{\Gsf}^\dagger$ therefore
casts the dual witness problem into Eq.~\eqref{eq:sdp} with
$f^\opt=1+\mathscr{A}_{\mathrm{R}}(\rho)$ for arbitrary symmetry
sector structures~\cite{SM}. Choosing $\zeta=\tau$ gives $c_\zeta=d_A$,
and the power-reweighted matrix takes the form
$X_q=V_{\Gsf}W^{[q]}V_{\Gsf}^\dagger$, yielding the witness
\begin{equation}\label{eq:asymmetry-witness}
 W^{[q]}\coloneqq
 \TCal_{\Gsf}(\rho^q)^{-1/2}\rho^q
 \TCal_{\Gsf}(\rho^q)^{-1/2}
\end{equation}
on $\HCal_B$, with objective
$f^{[q]}=\tr(\rho W^{[q]})$ and lower
estimate
$\mathscr{A}_{\mathrm{R}}^{[q]}(\rho)\coloneqq\max\{0,f^{[q]}-1\}$.
This witness is independent of the chosen Stinespring dilation and
satisfies
$\TCal_{\Gsf}(W^{[q]})=\proj_{\TCal_{\Gsf}(\rho^q)}\preceq I_B$,
making $\mathscr{A}_{\mathrm{R}}^{[q]}(\rho)$ a valid lower estimate of
$\mathscr{A}_{\mathrm{R}}(\rho)$ for every
representation. In particular, taking $\TCal_{\Gsf}$ to be complete
dephasing in a fixed basis of $\HCal_B$ yields power-reweighted witnesses
for the robustness of coherence with $d_A=d_B$~\cite{Napoli2016}.

All of these applications fulfill the pretty good performance at $q=1$ according to the BK bound,
and are collected in Table~\ref{tab:unification}. We next show how larger $q$ can lead to even better outcomes.

\begin{figure}[t]
\includegraphics[width=\columnwidth]{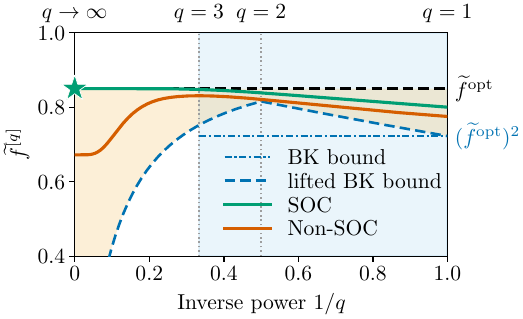}
\caption{Schematic of the lifted BK bound on normalized performance
$\ftilde^{[q]}$. The solid curves illustrate typical behavior for
a sectored orthogonal channel~(SOC) and a non-SOC channel under the same $\ftilde^{\opt}$.}
\label{fig:performance}
\end{figure}

\emph{Beyond the BK bound.---}Prior works have identified advantages of
the power-reweighted family for higher powers~\cite{BallesterWehnerWinter2008,Tyson2009,Tyson2010,ZhouChessaChitambarLeditzky2025}.
For the SDPs in Eq.~\eqref{eq:sdp}, we first establish a stronger lower bound valid for
every $q\ge1$ that recovers the BK bound at $q=1$.

\Needspace{5\baselineskip}
\begin{theorem}[Lifted BK bound]
\label{thm:lifted}
For every nonzero cost $C\succeq0$ and every $q\ge1$, choose any admissible
reference matrix $\zeta=\bigoplus_a\zeta_a$. Define the normalized objective values
$\ftilde^{[q]}\coloneqq f^{[q]}/c_\zeta$ and
$\ftilde^\opt\coloneqq f^\opt/c_\zeta$, and set
$r(q)\coloneqq\min\{q,2\}$. Then
$0<\ftilde^{[q]}\le\ftilde^\opt\le1$, and
\begin{equation}\label{eq:lifted}
 \ftilde^{[q]}
 \ge(\ftilde^\opt)^{\frac{q+1}{r(q)}}
 \left[\frac{2}{1+(\ftilde^\opt)^{1/2}}\right]^{2-2/r(q)}.
\end{equation}
The BK bound $\ftilde^{[q]}\ge(\ftilde^\opt)^2$ holds for
$q\in[1,3]$.
\end{theorem}

Theorem~\ref{thm:lifted} gives a single explicit performance guarantee
throughout the power-reweighted family. For a fixed normalized optimum
below one, the right-hand side of Eq.~\eqref{eq:lifted} increases from
the BK bound at $q=1$ to its maximum at $q=2$, and then decreases
for larger powers, as illustrated in Fig.~\ref{fig:performance}.
Thus $q=2$ supplies the strongest certificate within
this bound, while the best performing power for a particular task
depends on its cost and reference matrices. For $\ftilde^\opt=1-\eta$
($\eta\ll1$), $q=2$ reduces the leading guaranteed deficit from
$2\eta$~\cite{Zheng2024} to at most $5\eta/4$.

Across the broader family, the BK bound
$\ftilde^{[q]}\ge(\ftilde^\opt)^2$ holds universally on the
exact interval $q\in[1,3]$, and the lifted bound strictly improves on it
throughout $1<q\le3$ whenever $\ftilde^\opt<1$.
This improvement is universal for all schemes within this framework,
including all examples in Table~\ref{tab:unification}.
For every fixed real $q<1$ or $q>3$, the SM gives counterexamples
to the square bound in both channel recovery and pure-state
discrimination, each with a uniform reference~\cite{SM}.
For fixed $0<\ftilde^\opt<1$, the lifted bound vanishes as $q\to\infty$,
yet large $q$ can still be advantageous~\cite{ZhouChessaChitambarLeditzky2025}.
Generalizing the classical Bayesian model, whenever
$C=\sum_b C_b\otimes\kb{b}{b}_B$ in an orthonormal basis $\{\ket b\}$ of
$\HCal_B$, taking $Z\propto I$ achieves
$f^{[\infty]}=\sum_b\lambda_{\max}(C_b)=f^\opt$. We next present a practical
example in QEC with a different large-$q$ mechanism to gain further insight.

\emph{Maximum likelihood decoding.---}In QEC with encoder channel $\CCal$ and noise channel $\NCal$, usually one is interested in the channel
fidelity $\Fe(\tau,\RCal\circ\ECal)$, with $\ECal = \NCal\circ\CCal$, decoder $\RCal$, and a maximally mixed input state $\rho=\tau$.
Thus, we take $C=J_{\tau^2}(\ECal)$, $\zeta=\tau$, and $G_q=d_A C^q$.
The corresponding PPM decoder is the Petz map
$\PCal_{\tau,\widehat\FCal^{[q]}}=\PCal_{\tau,\ECal^{(q,\tau)}}$, where the
auxiliary CP map $\widehat\FCal^{[q]}=d_A^{-q}\ECal^{(q,\tau)}$ satisfies
$J(\widehat\FCal^{[q]})=G_q$.
A quantum channel $\ECal$ is an SOC if it has Kraus operators
$\sqrt{\mu_{s\ell}}E_{s\ell}$, where the positive weights
$\mu_{s\ell}$ sum to one and the partial isometries $E_{s\ell}:\HCal_A\to\HCal_B$ obey
\begin{equation}\label{eq:soc}
 \begin{gathered}
 E_{s\ell}^\dagger E_{s\ell}=\proj_A,\qquad
 E_{s\ell}E_{s\ell}^\dagger=\proj_s,\\
 \tr(E_{s\ell}^\dagger E_{tm})=d_A\delta_{st}\delta_{\ell m},\qquad
 \proj_s\proj_t=\delta_{st}\proj_s.
 \end{gathered}
\end{equation}
Here $s$ and $t$ label mutually orthogonal output sectors, while $\ell$ and $m$
label orthogonal logical operations within each sector.
All operators in a given sector have the same range, so
$U_{s\ell m}\coloneqq E_{s\ell}^\dagger E_{sm}$
is unitary on $\HCal_A$. We denote the set of all such quantum channels
by $\SOC$. The SOC conditions differ from those defining a nice error
basis~\cite{Knill1996}, while the sector geometry is closely related to the isoclinic subspace
framework for QEC~\cite{KribsMammarellaPereira2021,KribsPereiraTaank2025}.

The SOC class includes qubit and qudit stabilizer subspace codes under
stochastic Pauli (Weyl) noise~\cite{IyerPoulin2015,BermejoVegaVanDenNest2014}.
Grouping stabilizer equivalent errors gives orthogonal syndrome
spaces and Hilbert--Schmidt orthogonal logical Weyl
classes~\cite{Albert2020,BermejoVegaVanDenNest2014,SM}.
Ideal Gottesman--Kitaev--Preskill~(GKP) codes~\cite{GKP2001} with stochastic
displacement noise provide a continuous syndrome analogue, although continuous variable systems require more careful treatment in infinite dimensional Hilbert spaces.

Every SOC admits an MLD that measures the syndrome sector and reverses a most probable
logical error class within that sector.
Let $\RCal^{\MLD}$ denote the decoder that randomizes uniformly among
tied maximizing classes. On the populated output, its Kraus operators
are
\begin{equation}\label{eq:mld-kraus}
 R_{s\ell}^{\MLD}\coloneqq|\Lambda_s|^{-1/2}E_{s\ell}^\dagger,
 \qquad \ell\in\Lambda_s,
\end{equation}
where $\mu_{s,\max}\coloneqq\max_\ell\mu_{s\ell}$ is the maximum weight
in sector $s$ and
$\Lambda_s\coloneqq\{\ell:\mu_{s\ell}=\mu_{s,\max}\}$ is the set of
most likely logical error classes.
\begin{theorem}[Maximum likelihood limit]\label{thm:mld}
For $\ECal\in\SOC$ and $\zeta=\tau$, the channel fidelity $f^{[q]}$ is
nondecreasing as a function of $q>0$, with
\begin{equation}\label{eq:soc-fidelity}
 f^{[q]}=\sum_s\frac{\sum_\ell\mu_{s\ell}^{q+1}}
                         {\sum_\ell\mu_{s\ell}^{q}}
 \longrightarrow f^{\opt}=\sum_s \mu_{s,\max}
\end{equation}
as $q\to\infty$. The PPM decoder converges in diamond norm:
\begin{equation}\label{eq:mld-limit}
 \lim_{q\to\infty}\norm{\PCal_{\tau,\widehat\FCal^{[q]}}-\RCal^{\MLD}}_\diamond=0.
\end{equation}
\end{theorem}

Within each syndrome sector, larger $q$ concentrates the correction weights
onto the most likely logical error classes, while sector orthogonality
prevents competition between sectors, making the likelihood rule optimal
among all quantum decoders.
Outside the SOC class, the fidelity can still approach the optimum as
$q\to\infty$ or come close to it at a large finite $q$. For some non-SOC
channels, however, the fidelity decreases as $q$ increases from one~\cite{SM}.

\emph{Discussion.---}Our SDP framework unifies the power PGM, the PPM, and
broader quantum optimization tasks under a continuous lifted BK guarantee
for $q\ge1$, while SOC channels illuminate the large-$q$ reweighting
mechanism. This work indicates that it is possible to construct general ``pretty good''
channels, resource witnesses and other quantum operations as good approximate solutions 
with improved performance guarantees.

We are aware that the lifted BK bound in Eq.~\eqref{eq:lifted} is
generally not tight. Further refinements for $q>1$ are possible, which may cause the drift of the quadratic sweet-spot.
An improved bound requires more mathematical effort and may not have a simple
form. Determining the sharpest estimates is left to future work.
By clarifying how power reweighting
extends pretty good quantum operations, our framework supports further
applications in quantum information theory.

\begin{acknowledgments}
B.~L. thanks Victor V. Albert, Mohammad Alhejji, Aashish A. Clerk,
Yingkai Ouyang, and Yat Wong for helpful discussions.
We acknowledge support from the ARO (W911NF-23-1-0077),
ARO MURI (W911NF-21-1-0325), AFOSR MURI (FA9550-21-1-0209,
FA9550-23-1-0338), ONR MURI (N000142612102), DARPA (HR0011-24-9-0361),
and NSF (OSI-2326767, OSI-2426975, EEC-2550064, OSI-2553578, OSI-2553619).
This material is based upon work supported by the U.S. Department of
Energy, Office of Science, National Quantum Information Science Research
Centers and Advanced Scientific Computing Research~(ASCR) program under
contract number DE-AC02-06CH11357 as part of the InterQnet quantum
networking project.
We thank the ChatGPT for Academic Researchers program.
AI tools were used to assist with typesetting this manuscript and
developing the mathematical proofs.
\end{acknowledgments}
%

\clearpage
\onecolumngrid
\newpage
\setcounter{equation}{0}
\setcounter{figure}{0}
\setcounter{table}{0}
\setcounter{section}{0}
\setcounter{smtheorem}{0}
\renewcommand{\theequation}{S\arabic{equation}}
\renewcommand{\theHequation}{supp.\arabic{equation}}
\renewcommand{\thefigure}{S\arabic{figure}}
\renewcommand{\theHfigure}{supp.\arabic{figure}}
\renewcommand{\thetable}{S\arabic{table}}
\renewcommand{\theHtable}{supp.\arabic{table}}
\renewcommand{\thesection}{S\arabic{section}}
\renewcommand{\thesubsection}{\thesection.\Alph{subsection}}
\renewcommand{\thesubsubsection}{\thesubsection.\arabic{subsubsection}}
\makeatletter
\renewcommand{\p@subsection}{}
\renewcommand{\p@subsubsection}{}
\makeatother
\renewcommand{\theHsection}{supp.\arabic{section}}
\renewcommand{\theHsubsection}{\theHsection.\Alph{subsection}}
\renewcommand{\theHsubsubsection}{\theHsubsection.\arabic{subsubsection}}
\providecommand{\theHsmtheorem}{}
\renewcommand{\theHsmtheorem}{supp.\arabic{smtheorem}}
\setcounter{secnumdepth}{3}
\setcounter{tocdepth}{3}
\begingroup
\phantomsection\label{sec:supplement}
\begin{center}
{\large\bfseries Supplemental Material for\\
``Advancing the Theory of Pretty Good Quantum Operations''}\\[10pt]
Bikun Li, Guo Zheng, and Liang Jiang\\[2pt]
{\itshape Chicago Quantum Institute and Pritzker School of Molecular Engineering,\\
University of Chicago, Chicago, Illinois 60637, USA}
\end{center}
\addtocontents{toc}{\protect\smtocstart}
\supplementtableofcontents
\endgroup
\medskip

\section{Unified positive cost optimization and power reweighting}
\label{sec:sm:basic}
We use the main-text conventions and write
$\dket M\coloneqq\sum_a\ket a\otimes M\ket a$ for vectorization.
The decoder is trace preserving on the populated output
$\HCal_{B,+}=\supp(\tr_A C)$, as shown in
Sec.~\ref{sec:sm:basic-weighted}. On its orthogonal complement,
the choice of recovery does not affect the objective. For comparisons
in diamond norm, we use the same trace preserving completion there for
all decoders.

For an auxiliary CP map $\FCal$ with Choi matrix $M=J(\FCal)$, the
Petz map in Eq.~\eqref{eq:petz} at the reference $\tau$ satisfies
\begin{equation}\label{eq:sm:normalization}
 J(\PCal_{\tau,\FCal}^\dagger)
 =[I_A\otimes\left(\tr_A M\right)^{-1/2}]M
 [I_A\otimes\left(\tr_A M\right)^{-1/2}].
\end{equation}
The dual of Eq.~\eqref{eq:sdp} is
\begin{equation}\label{eq:sm:sdp}
 f^\opt=\min\{\tr\Gamma:\Gamma\succeq0,I_A\otimes\Gamma\succeq C\}.
\end{equation}
Strict feasibility gives equal primal and dual optima~\cite{Fletcher2007}.
One may also impose equality in the primal marginal constraint.
For feasible $X$, the matrix $X+\tau\otimes(I_B-\tr_A X)$ is positive
semidefinite, has marginal $I_B$, and achieves at least $\tr(CX)$ because
$C\succeq0$.

\subsection{Reference matrices and the power-reweighted construction}
\label{sec:sm:basic-weighted}
For the block references in Eq.~\eqref{eq:reference}, we omit zero cost
blocks and restrict each retained input space $\HCal_{A_a}$ to
$\supp(\zeta_a)$. Admissibility ensures that this restriction preserves
the support of $C_a$. Write $K\coloneqq Z^{1/2}$, so the seed in
Eq.~\eqref{eq:power-seed} is $G_q=K^{q-1}H^qK^{q-1}$.

For an explicit Kraus representation, factor the auxiliary matrix as
\begin{equation}\label{eq:sm:basic-decoder-kraus}
 \begin{aligned}
 G_q&=\sum_j\dket{L_{j,q}}\dbra{L_{j,q}},\qquad
 Y_q^2=\sum_jL_{j,q}L_{j,q}^\dagger,\\
 R_{j,q}&\coloneqq L_{j,q}^\dagger Y_q^{-1},\qquad
 \PCal_{\tau,\widehat\FCal^{[q]}}(\omega)=\sum_jR_{j,q}\omega R_{j,q}^\dagger.
 \end{aligned}
\end{equation}
Here $L_{j,q}:\HCal_{A}\to\HCal_{B}$ and
$\sum_jR_{j,q}^\dagger R_{j,q}=\proj_{Y_q}\preceq I_B$.

The construction with a uniform reference uses
$\zeta_a=I_{A_a}/\kappa$, where $\kappa\coloneqq\max_a\dim\HCal_{A_a}$. These
references are admissible even when the blocks have unequal dimensions,
and they give
\begin{equation}\label{eq:sm:basic-uniform-reference}
 Z=I_{AB}/\kappa,\qquad H=\kappa C,\qquad
 G_q=\kappa C^q,\qquad c_\zeta=\kappa\tr C.
\end{equation}
The common positive factor in $G_q$ cancels from its normalization.

For a single cost block, the cap can be minimized in closed form. We have
$c_\zeta=\tr(CZ^{-1})=\tr[(\tr_B C)\zeta^{-1}]$. Admissibility gives
$\supp(\tr_B C)\subseteq\supp(\zeta)$, and the Cauchy--Schwarz inequality
for the Hilbert--Schmidt inner product then yields
\begin{equation}\label{eq:sm:basic-cap-minimum}
 (\tr\sqrt{\tr_B C})^2
 =[\tr(\zeta^{1/2}\zeta^{-1/2}\sqrt{\tr_B C})]^2
 \le\tr(\zeta)\tr[(\tr_B C)\zeta^{-1}]\le c_\zeta,
\end{equation}
where the last step uses $\tr\zeta\le1$. The reference
$\zeta=\sqrt{\tr_B C}/\tr\sqrt{\tr_B C}$ attains this lower bound with
$c_\zeta=(\tr\sqrt{\tr_B C})^2$. It is admissible because a positive
semidefinite $C$ is supported in $\supp(\tr_B C)\otimes\HCal_B$.
Hence this choice minimizes the cap, as stated in the main text.

To verify normalization and feasibility, observe that
$\PCal_{\rho,\FCal}^\dagger(I_A)=\proj_{\FCal(\rho)}$ for every
completely positive map $\FCal$, and vectorizing its Kraus operators gives
Eq.~\eqref{eq:sm:normalization} because the factors $d_A^{-1/2}$ and
$d_A^{1/2}$ cancel. For any positive semidefinite $M$, a vector
$b\in\HCal_B$ lies in $\ker(\tr_A M)$ if and only if $M$ annihilates
$\HCal_A\otimes\operatorname{span}\{b\}$, since every nonnegative summand
of $\sum_a\bra{ab}M\ket{ab}$ must vanish. Because the block diagonal
operator $K$ is invertible on every retained summand and preserves
$\HCal_A\otimes\operatorname{span}\{b\}$, while every real power of $H$
shares the same kernel, applying this criterion to $G_q$ and $C=KHK$ shows
that $\ker(Y_q)=\ker(\tr_A C)$. Taking the partial trace in
Eq.~\eqref{eq:power-decoder} then yields
\begin{equation}\label{eq:sm:basic-support}
 \supp(Y_q)=\HCal_{B,+},\qquad
 \tr_A X_q=\proj_{Y_q}\preceq I_B,
\end{equation}
with equality on $\HCal_{B,+}$, establishing feasibility for every real
$q$.

\subsection{Relation to the RW iteration}
\label{sec:sm:basic-rw}
For a feasible $X$, one step of the RW iteration uses
\begin{equation}\label{eq:sm:basic-rw-update}
 \begin{aligned}
 G_X&\coloneqq\tr_A(CXC),\\
 \mathfrak T_C(X)&\coloneqq
 (I_A\otimes G_X^{-1/2})CXC(I_A\otimes G_X^{-1/2}).
 \end{aligned}
\end{equation}
This update has a nondecreasing objective~\cite{RW2005,Tyson2010}.
To obtain $X_2$ in one step, use the quadratic seed and initial point
\begin{equation}\label{eq:sm:basic-quadratic-rw}
 G_2=CZ^{-1}C,\qquad
 \gamma_\zeta\coloneqq\sum_a\tr(\zeta_a^{-1}),\qquad
 X^{(0)}\coloneqq\frac{Z^{-1}}{\gamma_\zeta}.
\end{equation}
Because $\tr_A X^{(0)}=I_B$ and the scalar $\gamma_\zeta^{-1}$ cancels from
the update, $\mathfrak T_C(X^{(0)})=X_2$. Further RW iterations
therefore preserve the power-reweighted decoder's lower bound by monotonicity.

Under the additional assumption $[Z,C]=0$, the seed reduces to
$G_q=Z^{-1}C^q$ for every $q\ge1$,
so one step of the RW iteration from $X^{(0)}\propto Z^{-1}C^{q-2}$ yields $X_q$.
Here powers are taken on $\supp(C)$, and a positive scalar rescaling
ensures $\tr_A X^{(0)}\preceq I_B$. Initialization with $C^{q-2}$ is
recovered when $Z\propto I$, as in the uniform reference construction.
Commutation alone does not remove the factor $Z^{-1}$.

When $Z$ and $C$ do not commute and $C$ is singular, the support of
$G_q$ can differ from $\supp(C)$, whereas every product $CX^{(0)}C$
is supported inside $\supp(C)$.
The quadratic identity in Eq.~\eqref{eq:sm:basic-quadratic-rw} holds for
every admissible reference.

On the populated output subspace $\HCal_{B,+}$, we have $\tr_A C\succ0$,
so global convergence of the RW iteration from a positive definite
initialization yields an optimizer $X^\opt$ satisfying
$\mathfrak T_C(X^\opt)=X^\opt$~\cite{RW2005,Ru2026}. Setting
$J(\FCal^\opt)\coloneqq CX^\opt C$ then matches
Eq.~\eqref{eq:sm:normalization}, giving
$X^\opt=J(\PCal_{\tau,\FCal^\opt}^\dagger)$.

\subsection{Examples}
\label{sec:sm:basic-examples}

\subsubsection{State discrimination}
\label{sec:sm:basic-discrimination}
For the discrimination cost $C=\bigoplus_i p_i\rho_i$ in
Table~\ref{tab:unification}, the scalar references $\zeta_i=1$ give
$c_\zeta=\tr C=1$.
Pinching a feasible $X$ onto these blocks preserves its objective and
output marginal, so it suffices to consider
$X=\sum_i\kb{i}{i}\otimes M_i$. The SDP therefore maximizes
$\sum_i p_i\tr(\rho_iM_i)$ over $M_i\succeq0$ with
$\sum_iM_i\preceq I_B$, where $I_B-\sum_iM_i$ is the inconclusive
effect.

The block structure gives
\begin{equation}\label{eq:sm:basic-pgm}
 G_q=C^q,\qquad Y_q=S_q^{1/2},\qquad
 X_q=\sum_i\kb{i}{i}\otimes M_i^{[q]},
\end{equation}
where $M_i^{[q]}$ are the effects of the power PGM in Eq.~\eqref{eq:genpgm}. At $q=2$,
this is also the first iterate of the
Je{\v z}ek--{\v R}eh{\'a}{\v c}ek--Fiur{\'a}{\v s}ek measurement
iteration from equal scalar multiples of the
identity~\cite{JezekRehacekFiurasek2002,Tyson2009,Tyson2010}.

For the preparation channel introduced in the main text,
$\ECal_{\mathrm{cq}}^\dagger(T)=\sum_i\tr(\rho_iT)\kb{i}{i}$ and
$\ECal_{\mathrm{cq}}(\sigma_{\mathrm{c}})=S_1$. Substitution into
Eq.~\eqref{eq:petz} gives the stated Petz representation of the
measurement channel of the ordinary PGM.

\subsubsection{Entanglement fidelity and its normalization}
\label{sec:sm:basic-fidelity}
Let $E_i$ and $R_j$ be Kraus operators of the channel $\ECal$ and the
recovery map $\RCal$, respectively. The entanglement fidelity at input
$\rho$ is
\begin{equation}\label{eq:sm:basic-entanglement-objective}
 \Fe(\rho,\RCal\circ\ECal)
 =\sum_{i,j}|\tr(\rho R_jE_i)|^2
 =\tr[J_{\rho^2}(\ECal)J(\RCal^\dagger)].
\end{equation}
Write $D_\rho\coloneqq(\sqrt{\rho})^{\T}\otimes I_B$ for the factor
in the dressed Choi matrix of Eq.~\eqref{eq:dressed}.
For the reference $\zeta=\rho^{\T}$, trace
preservation gives
\begin{equation}\label{eq:sm:basic-physical-reference}
 \begin{aligned}
 C&=J_{\rho^2}(\ECal)=D_\rho J_\rho(\ECal)D_\rho,\qquad
 Z=\rho^{\T}\otimes I_B=D_\rho^2,\\
 H&=J_\rho(\ECal),\qquad
 \tr_B C=(\rho^{\T})^2,\qquad c_\zeta=\tr H=1.
 \end{aligned}
\end{equation}
The reference and normalization cap remain fixed as the power varies, and
the seed in Eq.~\eqref{eq:power-seed} is
\begin{equation}\label{eq:sm:basic-physical-power}
 G_q=D_\rho^{q-1}[J_\rho(\ECal)]^qD_\rho^{q-1}.
\end{equation}
At unit power, $G_1=J_\rho(\ECal)$ and
$Y_1=\sqrt{\ECal(\rho)}$, so normalization gives the ordinary Petz map
$\PCal_{\rho,\ECal}$. At quadratic power,
\begin{equation}\label{eq:sm:basic-tyson-seed}
 G_2=D_\rho[J_\rho(\ECal)]^2D_\rho=CZ^{-1}C.
\end{equation}
To express the quadratic PPM directly in terms of $\ECal$, choose any
Kraus representation $\{E_i\}_i$ of this channel. The
reference-dependent quadratic auxiliary map in Proposition 45 of
Ref.~\cite{Tyson2010} is
\begin{equation}\label{eq:sm:basic-tyson-functional-calculus}
 \ECal^{(2,\rho)}(\omega)
 =\sum_{i,j}\tr(\rho E_i^\dagger E_j)E_i\omega E_j^\dagger.
\end{equation}
Vectorization and multiplication give
\begin{equation}\label{eq:sm:basic-tyson-kraus-matrix}
 \begin{aligned}
 G_2
 &=\sum_{i,j}\tr(\rho E_i^\dagger E_j)
 \dket{E_i\rho}\dbra{E_j\rho},\\
 Y_2^2&=\ECal^{(2,\rho)}(\rho^2).
 \end{aligned}
\end{equation}
The normalized recovery is therefore
\begin{equation}\label{eq:sm:basic-tyson-map}
 \PCal_{\rho^2,\ECal^{(2,\rho)}}(\omega)
 =\rho(\ECal^{(2,\rho)})^\dagger
 \bigl(Y_2^{-1}\omega Y_2^{-1}\bigr)\rho,
\end{equation}
which is the map in Theorem 44 of Ref.~\cite{Tyson2010}. More generally,
using the functional calculus of Definition 42 in that work, the general
representation in Eq.~\eqref{eq:petz-representation} connects to
Eq.~\eqref{eq:ppm} through
\begin{subequations}\label{eq:sm:basic-general-functional-calculus}
 \begin{align}
 J(\ECal^{(q,\rho)})
 &=D_\rho^{-1}[J_\rho(\ECal)]^qD_\rho^{-1},\label{eq:sm:basic-general-auxiliary-choi}\\
 G_q&=D_\rho^qJ(\ECal^{(q,\rho)})D_\rho^q,\label{eq:sm:basic-general-seed}\\
 \PCal_{\tau,\widehat\FCal^{[q]}}&=\PCal_{\rho^q,\ECal^{(q,\rho)}}.\label{eq:sm:basic-general-map}
 \end{align}
\end{subequations}
These formulas are understood on $\supp(\rho)$ and on
$\HCal_{B,+}=\supp(\ECal(\rho))$ when $\rho$ or $\ECal(\rho)$ is
singular.

For this reference, $c_\zeta=1$, so the normalized objective values
are the entanglement fidelities. Theorem~\ref{thm:sm:bk-range}
therefore applies directly, recovering the singleton-ensemble
Barnum--Knill bound at $q=1$~\cite{BK2002}.

For comparison, choosing the uniform reference $\zeta=\tau$ at the same
entanglement fidelity cost gives $c_\zeta=d_A\tr\rho^2$. Its unit power
decoder is $\PCal_{\rho^2,\ECal}$, and
its quadratic seed is proportional to $C^2$ and involves the
overlaps $\tr(\rho^2E_i^\dagger E_j)$ between Kraus operators. These generally differ from
those for $\zeta=\rho^{\T}$. When $\rho$ is maximally mixed,
the reference choices coincide and all auxiliary scalar factors cancel.

\subsubsection{Acceptance probability of a two-message verifier}
\label{sec:sm:verifier}
Every purification of the verifier's question state $\xi_B$ is obtained
from the canonical purification
$\sum_b\sqrt{\xi_B}^{\T}\ket{b}\otimes\ket{b}$ (memory first, question
second) by a partial isometry on the memory, which can be absorbed into
the accepting POVM effect $M_{\mathrm{acc}}$
($0\preceq M_{\mathrm{acc}}\preceq I_{AB}$, with the answer listed
first)~\cite{JainUpadhyayWatrous2009}. Taking the memory to be a copy of
$\HCal_B$, the joint state after the prover applies
$\SCal:\Lin(\HCal_B)\to\Lin(\HCal_A)$ to the question factor is directly
the dressed Choi matrix
$J_{\xi_B}(\SCal)=(\sqrt{\xi_B}^{\T}\otimes I_A)J(\SCal)(\sqrt{\xi_B}^{\T}\otimes I_A)$
on $\HCal_B\otimes\HCal_A$ from Eq.~\eqref{eq:dressed}. Because swapping
the two tensor factors maps $J(\SCal)$ to $J(\SCal^\dagger)^{\T}=X^{\T}$
for $X=J(\SCal^\dagger)$, the joint state of the answer and memory on
$\HCal_A\otimes\HCal_B$ is
$[(I_A\otimes\sqrt{\xi_B})X(I_A\otimes\sqrt{\xi_B})]^{\T}$. Pairing this
state with $M_{\mathrm{acc}}$ via $\tr(MN^{\T})=\tr(M^{\T}N)$ yields
\begin{equation}\label{eq:sm:verifier-acceptance}
 p_{\mathrm{acc}}(\SCal)=\tr(C_{\mathrm{ver}}X),
\end{equation}
with $C_{\mathrm{ver}}$ as in Eq.~\eqref{eq:verifier-cost}.

As $\SCal$ ranges over all quantum channels, $X$ ranges over all positive
semidefinite matrices with $\tr_A X=I_B$. Imposing this equality in
Eq.~\eqref{eq:sdp} leaves the optimum unchanged, as shown below
Eq.~\eqref{eq:sm:sdp}, giving $f^\opt=\max_{\SCal}p_{\mathrm{acc}}(\SCal)$.
Since $M_{\mathrm{acc}}^{\T}\preceq I_{AB}$ implies
$C_{\mathrm{ver}}\preceq I_A\otimes\xi_B$, choosing $\Gamma=\xi_B$ in
Eq.~\eqref{eq:sm:sdp} yields $f^\opt\le\tr\xi_B=1$. In the convention of
Ref.~\cite{JainUpadhyayWatrous2009}, where the output factor of a Choi
matrix is listed first, the verifier's interactive measurement operator is
$C_{\mathrm{ver}}^{\T}$, so transposition maps their SDP onto
Eq.~\eqref{eq:sdp}. Finally, by Eqs.~\eqref{eq:petz-representation}
and~\eqref{eq:sm:basic-support}, the map $\SCal^{[q]}$ in
Eq.~\eqref{eq:verifier-prover} has adjoint Choi matrix $X_q$ satisfying
$\tr_A X_q=\proj_{Y_q}$ and
$\supp(Y_q)=\HCal_{B,+}=\supp(\tr_A C_{\mathrm{ver}})$, so it achieves
$p_{\mathrm{acc}}(\SCal^{[q]})=f^{[q]}$. Appending Kraus operators
$\{K_m\}$ with $\sum_m K_m^\dagger K_m=I_B-\proj_{Y_q}$ completes
$\SCal^{[q]}$ to a trace preserving channel with adjoint Choi matrix
$X_q+\Delta X$, where $\Delta X\succeq0$ and
$\tr_A\Delta X=I_B-\proj_{Y_q}$ vanishes on $\HCal_{B,+}$. Thus
$C_{\mathrm{ver}}$ and $\Delta X$ are supported on orthogonal subspaces
of $\HCal_B$, giving $\tr(C_{\mathrm{ver}}\Delta X)=0$ and leaving the
acceptance probability unchanged.

\subsubsection{Robustness of asymmetry}
\label{sec:sm:asymmetry}
Let $g\mapsto U_g$ be any unitary representation of a finite or compact
group $\Gsf$ on $\HCal_B$, with group twirling map
$\TCal_{\Gsf}(M)=\int_{\Gsf}U_gMU_g^\dagger\mathrm{d}\mu(g)$
under the normalized Haar measure $\mu$~\cite{BartlettRudolphSpekkens2007}.
Invariance of the Haar measure under $g\mapsto g^{-1}$ makes
$\TCal_{\Gsf}$ self-adjoint, unital, trace preserving, and idempotent.
For any Kraus representation
$\TCal_{\Gsf}(M)=\sum_{a=1}^{d_A}K_aMK_a^\dagger$ and an orthonormal
basis $\{\ket{a}_A\}_{a=1}^{d_A}$ of $\HCal_A$, the Stinespring dilation
$V_{\Gsf}\coloneqq\sum_{a=1}^{d_A}\ket{a}_A\otimes K_a:\HCal_B\to\HCal_A\otimes\HCal_B$
obeys $V_{\Gsf}^\dagger V_{\Gsf}=I_B$,
$\tr_A(V_{\Gsf}MV_{\Gsf}^\dagger)=\TCal_{\Gsf}(M)$, and
$V_{\Gsf}^\dagger(I_A\otimes\Gamma)V_{\Gsf}=\TCal_{\Gsf}(\Gamma)$.
Moreover, its Choi matrix satisfies
$J(\TCal_{\Gsf})=\int_{\Gsf}\dket{U_g}\dbra{U_g}\mathrm{d}\mu(g)=\sum_{a=1}^{d_A}\dket{K_a}\dbra{K_a}$.
By continuity of the representation and full support of the Haar measure,
$\dbra{N}J(\TCal_{\Gsf})\dket{N}=\int_{\Gsf}|\dbraket{N}{U_g}|^2\mathrm{d}\mu(g)$
vanishes if and only if $\dbraket{N}{U_g}=0$ for all $g\in\Gsf$, so
$\supp(J(\TCal_{\Gsf}))=\operatorname{span}\{\dket{U_g}:g\in\Gsf\}$.
Each Kraus operator $K_a$ is therefore a linear combination of the
unitaries $\{U_g:g\in\Gsf\}$ and commutes with every
$\Gsf$-invariant operator $S$ on $\HCal_B$. Setting
$\proj_V\coloneqq V_{\Gsf}V_{\Gsf}^\dagger$, this gives
\begin{equation}\label{eq:sm:asymmetry-commutation}
 (I_A\otimes S)V_{\Gsf}=V_{\Gsf}S,
 \qquad [I_A\otimes S,\proj_V]=0
\end{equation}
whenever $\TCal_{\Gsf}(S)=S$.

We now show that setting
$C=V_{\Gsf}\rho V_{\Gsf}^\dagger$ casts
Eq.~\eqref{eq:asymmetry-definition} into Eq.~\eqref{eq:sdp} with
$f^\opt=1+\mathscr{A}_{\mathrm{R}}(\rho)$. Putting
$\tilde\sigma=(1+s)\sigma$ in the primal of
Eq.~\eqref{eq:asymmetry-definition} yields
$1+\mathscr{A}_{\mathrm{R}}(\rho)=\min\{\tr\tilde\sigma:\tilde\sigma\succeq0,\TCal_{\Gsf}(\tilde\sigma)=\tilde\sigma,\tilde\sigma\succeq\rho\}$~\cite{Piani2016Asymmetry},
whereas Eq.~\eqref{eq:sm:sdp} gives
$f^\opt=\min\{\tr\Gamma:\Gamma\succeq0,I_A\otimes\Gamma\succeq V_{\Gsf}\rho V_{\Gsf}^\dagger\}$.
If $\Gamma\succeq0$ is feasible in Eq.~\eqref{eq:sm:sdp}, then
$\tilde\sigma\coloneqq\TCal_{\Gsf}(\Gamma)=V_{\Gsf}^\dagger(I_A\otimes\Gamma)V_{\Gsf}\succeq\rho$
is positive semidefinite and $\Gsf$-invariant with
$\tr\tilde\sigma=\tr\Gamma$, proving
$1+\mathscr{A}_{\mathrm{R}}(\rho)\le f^\opt$. Conversely, if
$\tilde\sigma\succeq0$ is $\Gsf$-invariant and
$\tilde\sigma\succeq\rho$, then Eq.~\eqref{eq:sm:asymmetry-commutation}
gives
\begin{equation}\label{eq:sm:asymmetry-primal}
 \begin{aligned}
 I_A\otimes\tilde\sigma
 &=\proj_V(I_A\otimes\tilde\sigma)\proj_V
 +(I_{AB}-\proj_V)(I_A\otimes\tilde\sigma)(I_{AB}-\proj_V)\\
 &\succeq V_{\Gsf}\tilde\sigma V_{\Gsf}^\dagger
 \succeq V_{\Gsf}\rho V_{\Gsf}^\dagger,
 \end{aligned}
\end{equation}
so $\Gamma=\tilde\sigma$ is feasible in Eq.~\eqref{eq:sm:sdp} and
$f^\opt\le1+\mathscr{A}_{\mathrm{R}}(\rho)$.

Taking the uniform reference $\zeta=\tau$ gives $Z=I_{AB}/d_A$,
$H=d_A V_{\Gsf}\rho V_{\Gsf}^\dagger$, and normalization cap
$c_\zeta=\tr H=d_A$. Because $V_{\Gsf}$ is an isometry,
Eq.~\eqref{eq:power} yields
$G_q=d_A V_{\Gsf}\rho^q V_{\Gsf}^\dagger$ and
$Y_q^2=\tr_A G_q=d_A\TCal_{\Gsf}(\rho^q)$. Since
$\TCal_{\Gsf}(\rho^q)$ commutes with every $U_g$, its
Moore--Penrose inverse square root $\TCal_{\Gsf}(\rho^q)^{-1/2}$ is
also $\Gsf$-invariant. Applying Eq.~\eqref{eq:sm:asymmetry-commutation}
to $S=\TCal_{\Gsf}(\rho^q)^{-1/2}$ cancels $d_A$ in
Eq.~\eqref{eq:power-decoder} and yields
\begin{equation}\label{eq:sm:asymmetry-feasibility}
 X_q=V_{\Gsf}W^{[q]}V_{\Gsf}^\dagger,\qquad
 \TCal_{\Gsf}(W^{[q]})=\tr_A X_q=\proj_{\TCal_{\Gsf}(\rho^q)}
 \preceq I_B,
\end{equation}
using Eqs.~\eqref{eq:asymmetry-witness} and~\eqref{eq:sm:basic-support}.
Hence $W^{[q]}$ is independent of the chosen Stinespring dilation,
achieves $\tr(CX_q)=\tr(\rho W^{[q]})$, and yields the
lower estimate $\mathscr{A}_{\mathrm{R}}^{[q]}(\rho)\le\mathscr{A}_{\mathrm{R}}(\rho)$,
with Theorem~\ref{thm:sm:bk-range} governing $(1+\mathscr{A}_{\mathrm{R}}(\rho))/d_A$.
A minimal Stinespring dilation minimizes this uniform cap to the Choi
matrix rank $d_A=\rank[J(\TCal_{\Gsf})]=\sum_\lambda d_\lambda^2$,
where $d_\lambda$ are the dimensions of the distinct irreducible
representations appearing on $\HCal_B$. In particular, a single
irreducible sector of dimension $d_{\mathrm{irrep}}$ gives
$c_\zeta=d_{\mathrm{irrep}}^2$, whereas complete dephasing in a fixed
basis of $\HCal_B$ has $d_B$ one-dimensional sectors and recovers the
robustness of coherence with
$c_\zeta=d_B$~\cite{Piani2016Asymmetry,Napoli2016}.

\section{The BK bound and its sharp endpoints}
\label{sec:sm:bk}
We first restate Theorem~\ref{thm:lifted} of the main text, which this
section proves.

\begin{smtheorem}[Lifted BK bound]
\label{thm:sm:bk-range}
For every nonzero cost $C\succeq0$ and every $q\ge1$, choose any admissible
reference matrix $\zeta=\bigoplus_a\zeta_a$. Define the normalized objective values
$\ftilde^{[q]}\coloneqq f^{[q]}/c_\zeta$ and
$\ftilde^\opt\coloneqq f^\opt/c_\zeta$, and set
$r(q)\coloneqq\min\{q,2\}$. Then
$0<\ftilde^{[q]}\le\ftilde^\opt\le1$, and
\begin{equation}\label{eq:sm:bk-power-bound}
 \ftilde^{[q]}
 \ge(\ftilde^\opt)^{\frac{q+1}{r(q)}}
 \left[\frac{2}{1+(\ftilde^\opt)^{1/2}}\right]^{2-2/r(q)}.
\end{equation}
The BK bound $\ftilde^{[q]}\ge(\ftilde^\opt)^2$ holds for
$q\in[1,3]$.
\end{smtheorem}

Section~\ref{sec:sm:bk-proof} proves Theorem~\ref{thm:sm:bk-range}, and
Sec.~\ref{sec:sm:bk-sharpness} constructs counterexamples showing that
the interval $[1,3]$ is sharp for a universal square guarantee.

\subsection{Proof of the lifted BK bound}
\label{sec:sm:bk-proof}
With the support restrictions of Sec.~\ref{sec:sm:basic-weighted}, each
retained reference block $\zeta_a$ is positive definite (recall
$K=Z^{1/2}=\bigoplus_a(\zeta_a^{1/2}\otimes I_B)$), and we restrict
the output space to $\HCal_{B,+}$. Because pinching any feasible $X$ onto
the cost blocks $\HCal_{A_a}\otimes\HCal_B$ preserves positivity, the
partial trace, and $\tr(CX)$, it suffices to consider block diagonal $X$.
Diagonalizing each $\zeta_a$ with eigenvalues $z_i>0$ and factoring the
operator entries of $X_a$ in this eigenbasis as $(X_a)_{ij}=B_i^\dagger B_j$,
we have $B_i^\dagger B_i\preceq\tr_{A_a}X_a\preceq I_B$. For any vector
$v=(v_i)_i$ in that block, the Cauchy--Schwarz inequality gives
\begin{equation}\label{eq:sm:bk-weighted-contraction}
 \langle v|KX_aK|v\rangle
 =\left\lVert\sum_i\sqrt{z_i}B_iv_i\right\rVert^2
 \le\left(\sum_i z_i\right)\sum_i\lVert B_iv_i\rVert^2
 \le\sum_i\lVert v_i\rVert^2,
\end{equation}
using $\tr\zeta_a\le1$. Taking the direct sum over blocks establishes the
contraction bound
\begin{equation}\label{eq:sm:bk-block-bound}
 0\preceq KXK\preceq I_{AB}
\end{equation}
for every block diagonal feasible $X$. Since $C=KHK$ and $H\succeq0$, it
follows that $\tr(CX)=\tr(HKXK)\le\tr H=c_\zeta$, proving that $c_\zeta$
is indeed a cap on the optimum ($f^\opt\le c_\zeta$ and $\ftilde^\opt\le1$).

Multiplying $C$ by a positive scalar leaves $X_q$ unchanged and scales
$f^{[q]}$, $f^\opt$, and $c_\zeta$ by that same scalar. For the remainder
of this subsection, we therefore normalize $C$ so that
$\tr H=c_\zeta=1$, making $f^{[q]}=\ftilde^{[q]}$ and
$f^\opt=\ftilde^\opt$. From Eq.~\eqref{eq:sm:basic-support}, we have
$Y_q\succ0$ and $\tr_A X_q=I_B$ on $\HCal_{B,+}$. Because the feasible
set is closed and $\tr X\le\dim\HCal_B$, an optimizer $X^\opt$ exists.
Pinching it onto the input summands $\HCal_{A_a}$ and adding
$\omega_A\otimes(I_B-\tr_A X^\opt)$ for any density matrix $\omega_A$
supported in one retained summand preserves feasibility without lowering
the objective, so we may choose a block diagonal optimizer $X^\opt$ with
$\tr_A X^\opt=I_B$. Define the normalizer moment
\begin{equation}\label{eq:sm:bk-n}
 n_q\coloneqq\tr(Y_q^{2/q}),
\end{equation}
together with the positive definite operators $L_q$, $\Omega_q$ on the
retained joint space and the congruent optimizer $\widetilde X^\opt$,
\begin{subequations}\label{eq:sm:bk-trace-operators}
 \begin{align}
 L_q&\coloneqq K^q(I_A\otimes Y_q^{-1}),\label{eq:sm:bk-left-factor}\\
 \Omega_q&\coloneqq L_q^{-2/q}=K^{-2}(I_A\otimes Y_q^{2/q}),\label{eq:sm:bk-moment-operator}\\
 \widetilde X^\opt&\coloneqq KX^\opt K.\label{eq:sm:bk-congruent-optimizer}
 \end{align}
\end{subequations}
For $q>1$, the normalizer moment $n_q$ has an entropic meaning on a
purifying system. Suppose that $\zeta$ is a single reference of unit
trace. Let $\ket{\Psi}$ purify $H$ on an auxiliary system $A'$, with
reduced states $\varrho_{AA'}$ and $\varrho_{A'}$. The vector
$(\zeta^{(q-1)/2}\otimes I_B\otimes\varrho_{A'}^{(q-1)/2})\ket{\Psi}$ has
the reduced operator $Y_q^2$ on $\HCal_B$, because powers of
$\varrho_{A'}$ acting on $\ket{\Psi}$ can be exchanged for the same powers
of $H$. Comparing the nonzero spectra of its two reductions gives
\begin{equation}\label{eq:sm:bk-n-renyi}
 \log n_q=-\frac{q-1}{q}
 \widetilde D_{1/q}(\varrho_{AA'}\Vert\zeta\otimes\varrho_{A'}),
\end{equation}
where $\widetilde D_\alpha$ denotes the sandwiched R\'enyi relative
entropy of order $\alpha$~\cite{Beigi2013}. The moment $n_q$ therefore
evaluates this relative entropy at the marginal $\varrho_{A'}$ instead of
minimizing it over all states on the purifying system. This substitution
increases the relative entropy by at most the factor
$q$~\cite{SchmittJiangLiRenes2026}.

Because $K$ commutes with every output operator, substituting
Eqs.~\eqref{eq:sm:bk-trace-operators} into Eq.~\eqref{eq:power-decoder}
gives $KX_qK=L_qH^qL_q$. Moreover, trace cyclicity and $\tr_A X^\opt=I_B$
give
$\tr(\Omega_q\widetilde X^\opt)=\tr[(I_A\otimes Y_q^{2/q})X^\opt]=\tr(Y_q^{2/q}\tr_A X^\opt)=n_q$,
while substituting $Y_q^2=\tr_A(K^{q-1}H^qK^{q-1})$ into
$n_q=\tr[Y_q^2Y_q^{-2(q-1)/q}]$ gives $n_q=\tr(H^qL_q^{2(q-1)/q})$.
Applying Eq.~\eqref{eq:sm:bk-block-bound} to $X_q$ and $X^\opt$ and
noting that $L_qH^{q/2}$ and its adjoint share the same operator norm
therefore yields the trace identities and contraction bounds
\begin{subequations}\label{eq:sm:bk-trace-relations}
 \begin{gather}
 \ftilde^{[q]}=\tr(HL_qH^qL_q),\qquad
 \ftilde^\opt=\tr(H\widetilde X^\opt),\label{eq:sm:bk-trace-objectives}\\
 L_qH^qL_q\preceq I_{AB},\qquad
 H^{q/2}L_q^2H^{q/2}\preceq I_{AB},\qquad
 0\preceq\widetilde X^\opt\preceq I_{AB},\label{eq:sm:bk-contractions}\\
 n_q=\tr(\Omega_q\widetilde X^\opt)=\tr(H^qL_q^{2(q-1)/q}).\label{eq:sm:bk-trace-nq}
 \end{gather}
\end{subequations}
In an eigenbasis of $H$, the diagonal entries of $HL_qH^qL_q$ have a
strictly positive sum because $H\ne0$ and $L_q\succ0$, so
$0<\ftilde^{[q]}\le\ftilde^\opt\le1$. All powers of $L_q$ in
Eqs.~\eqref{eq:sm:bk-trace-relations} act on the full retained joint
space.

To link the normalizer moment $n_q$ to the trace
$m_q\coloneqq\tr(H^{1+q/2}L_q)>0$, we establish a short moment estimate.

\begin{smlemma}[Normalizer moments]\label{lem:sm:bk-moments}
With $\tr H=1$ and $m_q\coloneqq\tr(H^{1+q/2}L_q)>0$, the normalizer
moment satisfies
\begin{equation}\label{eq:sm:bk-moment-comparison}
 n_q^{q/2}\le
 \begin{cases}
 m_q^{q-1},&1\le q\le2,\\
 m_q,&q\ge2.
 \end{cases}
\end{equation}
\end{smlemma}
\begin{proof}
Let $h_i>0$ be the eigenvalues of $H$ on its support and let $\ell_j>0$
be the eigenvalues of $L_q$ on the full retained space, with respective
orthonormal eigenvectors $u_i$ and $v_j$. Assign $(i,j)$ the probability
$h_i|\langle u_i,v_j\rangle|^2$, which sums to $\tr H=1$, and set
$z_{ij}\coloneqq h_i^{q/2}\ell_j$. Writing $\mathbb{E}[\cdot]$ for
expectation under this distribution,
Eqs.~\eqref{eq:sm:bk-contractions} and~\eqref{eq:sm:bk-trace-nq}
give
\begin{equation}\label{eq:sm:bk-scalar-moments}
 \mathbb{E}[z]=m_q,\qquad
 \mathbb{E}[z^{2(q-1)/q}]=n_q,\qquad
 \mathbb{E}[z^2]=\tr(H^{q+1}L_q^2)\le\tr H=1.
\end{equation}
For $0\le t\le1$, Jensen's inequality gives
$\mathbb{E}[z^t]\le m_q^t$. For $1\le t<2$, the scalar H\"older
inequality gives
$\mathbb{E}[z^t]\le(\mathbb{E}[z])^{2-t}(\mathbb{E}[z^2])^{t-1}\le m_q^{2-t}$.
Setting $t=2(q-1)/q$ and raising both sides to the power $q/2$ proves
both ranges.
\end{proof}

We now assemble the proof of Theorem~\ref{thm:sm:bk-range}. Throughout,
we use the Hilbert--Schmidt inner product
$\langle U,V\rangle=\tr(U^\dagger V)$ and the Schatten norms
$\lVert T\rVert_t=(\tr|T|^t)^{1/t}$, where $|T|=(T^\dagger T)^{1/2}$ and
$1\le t<\infty$. Both power ranges rely on the Buzano
inequality~\cite{Dragomir2016}
\begin{equation}\label{eq:sm:bk-buzano}
 |\langle x,z\rangle\langle z,y\rangle|
 \le\frac{\lVert x\rVert\lVert y\rVert+|\langle x,y\rangle|}{2}
 \lVert z\rVert^2,
\end{equation}
applied to matrices under the Hilbert--Schmidt inner product.

\begin{proof}[Proof of Theorem~\ref{thm:sm:bk-range}]
With the normalization $\tr H=c_\zeta=1$ established above, we already
have $0<\ftilde^{[q]}\le\ftilde^\opt\le1$, and positive rescaling of $C$
restores arbitrary $c_\zeta$. We treat the two power ranges separately.

\emph{Case 1: $1\le q\le2$.} Put $p\coloneqq q/2\in[1/2,1]$ and
$E^\opt\coloneqq H^{1/2}\widetilde X^\opt H^{1/2}$, so that
$0\preceq E^\opt\preceq H$ and $\tr E^\opt=\ftilde^\opt$. Define
$R\coloneqq(E^\opt)^{p/2}H^{1/2}\Omega_q^{-p/2}$ and
$T\coloneqq\Omega_q^{p/2}(\widetilde X^\opt)^{1/2}$. Because
$(\widetilde X^\opt)^{1/2}$ is a contraction and $t\mapsto t^{1/p}$ is
convex on $[0,\infty)$ with value zero at the origin, the contractive
trace Jensen inequality~\cite{HansenPedersen2003} gives
\begin{equation}\label{eq:sm:bk-trace-jensen}
 \lVert T\rVert_{2/p}^2
 =\left\{\tr\bigl[\bigl((\widetilde X^\opt)^{1/2}\Omega_q^p
 (\widetilde X^\opt)^{1/2}\bigr)^{1/p}\bigr]\right\}^p
 \le[\tr(\Omega_q\widetilde X^\opt)]^p=n_q^p.
\end{equation}
Since $(RT)(RT)^\dagger=(E^\opt)^{1+p}$ and $\Omega_q^{-p}=L_q$, the
Schatten H\"older inequality~\cite{Beigi2013} yields
\begin{equation}\label{eq:sm:bk-lower-resource}
 (\ftilde^\opt)^{1+p}
 =\lVert RT\rVert_{2/(1+p)}^2
 \le\lVert R\rVert_2^2\lVert T\rVert_{2/p}^2
 \le n_q^p\tr\bigl[(E^\opt)^pH^{1/2}L_qH^{1/2}\bigr].
\end{equation}
Only $\Omega_q\succ0$ is inverted, so this estimate covers singular $H$
and $\widetilde X^\opt$ as well.

To bound the remaining trace, restrict to $\supp(H)$ so that $H\succ0$,
and put $N\coloneqq H^{1/2}$ and $M\coloneqq H^{q/2}L_qH^{1/2}$. Then
$\lVert N\rVert_2=1$, $\lVert M\rVert_2^2=\ftilde^{[q]}$, and
$\langle N,M\rangle=m_q>0$. Assume first that $E^\opt\succ0$ on
$\supp(H)$ and $1<q<2$, and set $\theta\coloneqq q-1\in(0,1)$. On the
strip $0\le\mathrm{Re}(z)\le1$, consider the bounded analytic function
\begin{equation}\label{eq:sm:bk-lower-analytic}
 \varphi(z)\coloneqq\tr\bigl[(E^\opt)^{(1+z)/2}H^{-z/2}M\bigr],
\end{equation}
which satisfies
$\varphi(\theta)=\tr[(E^\opt)^pH^{1/2}L_qH^{1/2}]$. Because imaginary
powers of positive definite matrices are unitary, the Cauchy--Schwarz
inequality on the left boundary gives
$|\varphi(it)|\le\sqrt{\ftilde^\opt}\lVert M\rVert_2=\sqrt{\ftilde^\opt\ftilde^{[q]}}$.
On the right boundary, $\varphi(1+it)=\langle U_t,M\rangle$ with
$U_t\coloneqq H^{-1/2+it/2}(E^\opt)^{1-it/2}$. Using
$E^\opt H^{-1}E^\opt\preceq E^\opt$ (from $E^\opt\preceq H$) and
$|\tr(E^\opt U)|\le\tr E^\opt$ for unitary $U$, we obtain
\begin{subequations}\label{eq:sm:bk-lower-boundary}
 \begin{align}
 \lVert U_t\rVert_2^2&=\tr[(E^\opt)^2H^{-1}]\le\ftilde^\opt,\label{eq:sm:bk-lower-boundary-norm}\\
 |\langle N,U_t\rangle|&=|\tr[H^{it/2}(E^\opt)^{1-it/2}]|\le\ftilde^\opt.\label{eq:sm:bk-lower-boundary-inner}
 \end{align}
\end{subequations}
Applying the Buzano inequality in Eq.~\eqref{eq:sm:bk-buzano} with
$x=U_t$, $y=N$, and $z=M$ gives
$|\varphi(1+it)|m_q\le\frac{1}{2}(\sqrt{\ftilde^\opt}+\ftilde^\opt)\ftilde^{[q]}$.
By the Hadamard three-lines theorem~\cite{Beigi2013} applied to
$\varphi(z)m_q^z$ at $z=\theta=q-1$,
\begin{equation}\label{eq:sm:bk-lower-strip}
 \varphi(\theta)m_q^{q-1}
 \le\bigl(\sqrt{\ftilde^\opt\ftilde^{[q]}}\bigr)^{2-q}
 \left(\frac{\sqrt{\ftilde^\opt}+\ftilde^\opt}{2}
 \ftilde^{[q]}\right)^{q-1}
 =\sqrt{\ftilde^\opt}
 \left(\frac{1+\sqrt{\ftilde^\opt}}{2}\right)^{q-1}
 (\ftilde^{[q]})^{q/2}.
\end{equation}
At $q=1$ and $q=2$, the same inequality follows directly from the left
and right boundary estimates, respectively. For singular $E^\opt$ on
$\supp(H)$, replacing $E^\opt$ by $E_\delta=(1-\delta)E^\opt+\delta H$
($0<\delta<1$) and taking $\delta\to0^+$ preserves
Eq.~\eqref{eq:sm:bk-lower-strip} by continuity of positive powers.
Combining Eqs.~\eqref{eq:sm:bk-lower-resource} and~\eqref{eq:sm:bk-lower-strip}
with Lemma~\ref{lem:sm:bk-moments} ($n_q^{q/2}\le m_q^{q-1}$) gives
\begin{equation}\label{eq:sm:bk-lower-final}
 (\ftilde^\opt)^{1+q/2}
 \le\sqrt{\ftilde^\opt}
 \left(\frac{1+\sqrt{\ftilde^\opt}}{2}\right)^{q-1}
 (\ftilde^{[q]})^{q/2}.
\end{equation}
Dividing by $\sqrt{\ftilde^\opt}$ and raising to the power $2/q$ proves
Eq.~\eqref{eq:sm:bk-power-bound} for $1\le q\le2$, including at $q=1$
where $n_1=1$.

\emph{Case 2: $q\ge2$.} Put $r\coloneqq q/2\ge1$, $N\coloneqq H^{1/2}$,
and $M\coloneqq H^{r/2}L_qH^{(r+1)/2}$. Then $\lVert N\rVert_2=1$,
$\langle N,M\rangle=m_q>0$, and $\lVert M\rVert_2^2\le\ftilde^{[q]}$.
Indeed, diagonalizing $H$ on $\supp(H)$ with eigenvalues $h_i>0$ and
using Hermiticity of $L_q$ yields
\begin{equation}\label{eq:sm:bk-upper-trace-difference}
 \ftilde^{[q]}-\lVert M\rVert_2^2
 =\frac{1}{2}\sum_{i,j}h_ih_j
 (h_i^r-h_j^r)(h_i^{r-1}-h_j^{r-1})|(L_q)_{ij}|^2\ge0,
\end{equation}
since both scalar powers are nondecreasing for $r\ge1$ and terms
involving $\ker(H)$ vanish in both traces.

On the strip $0\le\mathrm{Re}(z)\le1$, consider the matrix function
\begin{equation}\label{eq:sm:bk-upper-analytic}
 \psi(z)\coloneqq L_q^{z/2}H^{(1+rz)/2}(\widetilde X^\opt)^{1/2},
\end{equation}
where complex powers of $H$ act on $\supp(H)$ and vanish on $\ker(H)$,
keeping $\psi$ bounded and analytic even when $H$ is singular. For real
$t$, define the unitary $V_t$ to be $H^{irt/2}$ on $\supp(H)$ and the
identity on $\ker(H)$. Because $V_t$ commutes with $H$, the operator
$\widetilde X_t^\opt\coloneqq V_t\widetilde X^\opt V_t^\dagger$ obeys
$0\preceq\widetilde X_t^\opt\preceq I_{AB}$ and
$\tr(H\widetilde X_t^\opt)=\ftilde^\opt$. Unitary
invariance on the two boundaries gives
\begin{equation}\label{eq:sm:bk-upper-boundary}
 \lVert\psi(it)\rVert_2^2=\ftilde^\opt,\qquad
 \lVert\psi(1+it)\rVert_2^2
 =\tr[\widetilde X_t^\opt H^{(r+1)/2}L_qH^{(r+1)/2}]
 =\langle N\widetilde X_t^\opt,M\rangle.
\end{equation}
Since $\lVert N\widetilde X_t^\opt\rVert_2^2=\tr[H(\widetilde X_t^\opt)^2]\le\ftilde^\opt$ and
$\langle N\widetilde X_t^\opt,N\rangle=\ftilde^\opt$, the Buzano inequality in
Eq.~\eqref{eq:sm:bk-buzano} gives
$m_q\lVert\psi(1+it)\rVert_2^2\le\frac{1}{2}(\sqrt{\ftilde^\opt}+\ftilde^\opt)\lVert M\rVert_2^2\le\frac{1}{2}(\sqrt{\ftilde^\opt}+\ftilde^\opt)\ftilde^{[q]}$.
For $r>1$, the matrix three-lines theorem~\cite{Beigi2013} with both
endpoint Schatten exponents equal to $2$, evaluated at $z=1/r$, yields
\begin{equation}\label{eq:sm:bk-upper-strip}
 \lVert L_q^{1/(2r)}H(\widetilde X^\opt)^{1/2}\rVert_2^2
 \le(\ftilde^\opt)^{1-1/r}
 \left(\frac{(\sqrt{\ftilde^\opt}+\ftilde^\opt)\ftilde^{[q]}}
 {2m_q}\right)^{1/r},
\end{equation}
which at $r=1$ is simply the right boundary estimate. Applying the
Cauchy--Schwarz inequality to
$\ftilde^\opt=\langle L_q^{-1/(2r)}(\widetilde X^\opt)^{1/2},L_q^{1/(2r)}H(\widetilde X^\opt)^{1/2}\rangle$
and noting that
$\lVert L_q^{-1/(2r)}(\widetilde X^\opt)^{1/2}\rVert_2^2=\tr(\Omega_q\widetilde X^\opt)=n_q$
gives
\begin{equation}\label{eq:sm:bk-upper-budget}
 (\ftilde^\opt)^2
 \le n_q\lVert L_q^{1/(2r)}H(\widetilde X^\opt)^{1/2}\rVert_2^2
 \le n_q(\ftilde^\opt)^{1-1/r}
 \left(\frac{(\sqrt{\ftilde^\opt}+\ftilde^\opt)\ftilde^{[q]}}
 {2m_q}\right)^{1/r}.
\end{equation}
Dividing by $(\ftilde^\opt)^{1-1/r}$, raising to the power $r=q/2$,
multiplying by $m_q$, and applying Lemma~\ref{lem:sm:bk-moments}
($n_q^{q/2}\le m_q$) yields
\begin{equation}\label{eq:sm:bk-upper-final}
 (\ftilde^\opt)^{1+q/2}m_q
 \le\frac{\sqrt{\ftilde^\opt}+\ftilde^\opt}{2}n_q^{q/2}\ftilde^{[q]}
 \le\frac{\sqrt{\ftilde^\opt}+\ftilde^\opt}{2}m_q\ftilde^{[q]}.
\end{equation}
Dividing by $m_q\sqrt{\ftilde^\opt}>0$ proves
Eq.~\eqref{eq:sm:bk-power-bound} for $q\ge2$. Both ranges agree at $q=2$
and include $\ftilde^\opt=1$.

Finally, on $[1,3]$ the exponent $(q+1)/r(q)$ in
Eq.~\eqref{eq:sm:bk-power-bound} is at most $2$, while the factor
$2/[1+(\ftilde^\opt)^{1/2}]$ is at least $1$ and strictly exceeds $1$
whenever $1<q\le3$ and $\ftilde^\opt<1$. This establishes the BK bound on
$[1,3]$ and its strict improvement on $(1,3]$. The counterexamples in
Sec.~\ref{sec:sm:bk-sharpness} exclude every real power outside $[1,3]$
for both channel fidelity and pure-state discrimination with admissible
uniform references, showing that the interval $[1,3]$ is exact.
\end{proof}

\subsection{Sharpness of the universal range}
\label{sec:sm:bk-sharpness}
The interval $[1,3]$ concerns a guarantee uniform over all normalized
optima. For a fixed $0<\ftilde^\opt<1$, the lifted bound can remain
above the square bound beyond $q=3$. To establish the sharp universal
range, we now construct counterexamples whose normalized optima vary
with the construction parameter. In each case, take
$\zeta_a=I_{A_a}/\kappa$ and set $T\coloneqq\tr C$, so $c_\zeta=\kappa T$.
With these references, Eq.~\eqref{eq:sm:basic-uniform-reference} gives
a seed proportional to $C^q$.

\begin{smlemma}[Failure below one]\label{lem:sm:bk-below-one}
For each real $q<1$, the inequality $f^{[q]}\geq(f^\opt)^2/c_\zeta$ fails
for some qubit channel-fidelity cost and for some pure-state discrimination cost, both with $c_\zeta=1$.
\end{smlemma}

\begin{proof}
We demonstrate this failure by an SOC example. Let $\sigma_x:=\kb{0}{1}+\kb{1}{0}$. For an integer
$k\geq3$, consider
\begin{equation}
 \ECal_k(\rho)\coloneqq \frac{k}{k+1}\rho+
 \frac1{k+1}\sigma_x\rho \sigma_x^\dagger.
 \label{eq:sm:bk-bit-flip}
\end{equation}
Use $C=J(\ECal_k)/4$ as a single cost sub-block, so
$\kappa=2$, $T=1/2$, and $c_\zeta=1$. The two Choi matrix eigenvectors with nonzero eigenvalues are
orthogonal Bell vectors with identical maximally mixed marginals. Direct normalization therefore gives
\begin{equation}
 f^\opt=\frac{k}{k+1},
 \qquad
 f^{[q]}=\frac{k^{q+1}+1}{(k+1)(k^q+1)}.
 \label{eq:sm:bk-bit-flip-values}
\end{equation}
For completeness, $C\preceq f^\opt I_4/2$ and every feasible $X$ satisfies
$\tr X\leq2$, so its objective is at most $f^\opt$. Identity recovery
attains that value. The exact gap is
\begin{equation}
 f^{[q]}-(f^\opt)^2=
 \frac{k^{q+1}-k^2+k+1}{(k+1)^2(k^q+1)}.
 \label{eq:sm:bk-bit-flip-gap}
\end{equation}
For any fixed real $q<1$, the numerator is negative for all sufficiently
large $k$. Thus no power below one satisfies the universal
square guarantee.

The same gap occurs in state discrimination. Take two identical pure
states with priors $k/(k+1)$ and $1/(k+1)$. The power PGM
guesses a label with probability proportional to the $q$th power of its
prior, whereas the optimal measurement always chooses the more likely
label. Hence the same values and gap hold, now with one-dimensional
label spaces and $\kappa=T=c_\zeta=1$.
\end{proof}

To construct counterexamples above the upper endpoint, we use
overlapping output ranges.

\begin{smlemma}[Failure above three]\label{lem:sm:bk-above-three}
For each $q>3$, the inequality $f^{[q]}\geq(f^\opt)^2/c_\zeta$ fails for
some finite-dimensional channel-fidelity cost and for some pure-state discrimination cost, both with $c_\zeta=1$.
\end{smlemma}

\begin{proof}
Fix $q>3$. For an integer $k\geq2$, set
\begin{equation}
 a=k^{-1},\qquad m=2k^2,\qquad N=2k^4=m^2/2,
 \qquad t=a^2/4.
 \label{eq:sm:bk-upper-parameters}
\end{equation}
Partition a Weyl unitary basis on $\mathbb C^m$ into two sets of
$N$ operators, denoted $U_i$ and $V_i$. Thus all $2N$ operators
are pairwise orthogonal in the Hilbert--Schmidt inner product.
Let $e_0,e_1,\ldots,e_N$ be an orthonormal syndrome basis and put
$h_i\coloneqq \sqrt{1-t}e_0+\sqrt t e_i$. Define the channel
$\ECal:\Lin(\mathbb C^m)\to\Lin(\mathbb C^m\otimes\mathbb C^{N+1})$
by the Kraus operators
\begin{equation}
 E_{h,i}\coloneqq \sqrt{(1-a)/N}(U_i\otimes h_i),
 \qquad E_{g,i}\coloneqq \sqrt{a/N}(V_i\otimes e_i),
 \label{eq:sm:bk-upper-channel}
\end{equation}
where $(U\otimes h)v\coloneqq Uv\otimes h$. Their adjoint products sum
to $I_m$, so the channel is trace preserving. Take
$C=J(\ECal)/m^2$ as a single cost sub-block, giving
$\kappa=m$, $T=1/m$, and $c_\zeta=1$. Measuring the syndrome in the basis
$\{e_i\}_{i=0}^N$ and applying $V_i^\dagger$ for outcome $e_i$ with
$i\ge1$ recovers every branch with Kraus operator $E_{g,i}$. These
branches have total probability $a$, so $f^\opt\geq a$.

For every Weyl operator $U$, define the normalized Bell vector
$\phi_U\coloneqq m^{-1/2}\sum_x\ket{x}\otimes U\ket{x}$. Orthogonality
of these logical Bell vectors makes the corresponding Choi matrix
eigenvectors orthogonal even though their syndrome components overlap. Put
\begin{equation}
 r\coloneqq \frac a{1-a},\qquad z\coloneqq r^q,
 \qquad
 \Omega\coloneqq \sum_i\ket{h_i}\!\bra{h_i}+
 z\sum_i\ket{e_i}\!\bra{e_i}.
 \label{eq:sm:bk-upper-normalizer}
\end{equation}
Taking the spectral power in this orthogonal decomposition and then
tracing out the input gives
\begin{equation}
 f^{[q]}=(1-a)\alpha_h^2+az\alpha_e^2,
 \qquad \alpha_h\coloneqq \bra{h_i}\Omega^{-1/2}\ket{h_i},
 \qquad \alpha_e\coloneqq \bra{e_i}\Omega^{-1/2}\ket{e_i}.
 \label{eq:sm:bk-upper-fidelity}
\end{equation}
Permutation symmetry makes $\alpha_h$ and $\alpha_e$ independent of $i$. We will use
the elementary estimates
\begin{equation}
 \alpha_h\leq N^{-1/2}+\sqrt t,
 \qquad \alpha_e\leq(N\sqrt z)^{-1}+t^{-1/2}.
 \label{eq:sm:bk-upper-estimates}
\end{equation}
For $u\coloneqq N^{-1/2}\sum_i e_i$ and $A_0\coloneqq N(1-t)$, the matrix of
$\Omega$ on $\mathrm{span}\{e_0,u\}$ is
\begin{equation}
 M_0\coloneqq \begin{pmatrix}A_0&\sqrt{A_0t}\\
 \sqrt{A_0t}&t+z\end{pmatrix},
 \qquad
 M_0^{-1/2}=\frac1{s\ell_0}
 \begin{pmatrix}t+z+s&-\sqrt{A_0t}\\
 -\sqrt{A_0t}&A_0+s\end{pmatrix},
 \label{eq:sm:bk-two-dimensional-block}
\end{equation}
where $s\coloneqq \sqrt{A_0z}$ and $\ell_0\coloneqq \sqrt{A_0+t+z+2s}$.
On its orthogonal complement in the syndrome space,
$\Omega=(t+z)I$. Hence
\begin{equation}
 \begin{aligned}
 \alpha_h&=\frac{A_0+t+s}{N\ell_0}+\left(1-\frac1N\right)\frac t{\sqrt{t+z}},\\
 \alpha_e&=\frac{A_0+s}{Ns\ell_0}+\left(1-\frac1N\right)\frac1{\sqrt{t+z}}.
 \end{aligned}
 \label{eq:sm:bk-upper-expectations}
\end{equation}
The inequalities
$(A_0+t+s)/\ell_0\leq\sqrt{A_0+t}\leq\sqrt N$ and
$(A_0+s)/(s\ell_0)\leq z^{-1/2}$ follow by squaring. They prove
Eq.~\eqref{eq:sm:bk-upper-estimates}, which in turn yields
\begin{equation}
 f^{[q]}\leq(1-a)\left[
 (N^{-1/2}+\sqrt t)^2+
 \left(\frac{\sqrt r}{N}+\frac{r^{(q+1)/2}}{\sqrt t}\right)^2\right].
 \label{eq:sm:bk-upper-bound}
\end{equation}
Dividing this bound by $a^2$ and substituting
Eq.~\eqref{eq:sm:bk-upper-parameters} gives
\begin{equation}
 (1-k^{-1})\left[
 \left(\frac1{\sqrt2 k}+\frac12\right)^2+
 \left(\frac{\sqrt r}{2k^3}+2k^2r^{(q+1)/2}\right)^2\right]
 \longrightarrow\frac14.
 \label{eq:sm:bk-upper-limit}
\end{equation}
Here the limit uses $q>3$ and $r=1/(k-1)$. Hence, for every sufficiently
large integer $k$, $f^{[q]}<a^2\leq(f^\opt)^2$, which supplies a
finite-dimensional channel counterexample for each prescribed $q>3$.

An analogous pure-state ensemble gives the state-discrimination
counterexample. Use the $2N$ pure states $\kb{h_i}{h_i}$ and
$\kb{e_i}{e_i}$ with priors $(1-a)/N$ and $a/N$, respectively.
The sum of their powered ensemble operators is $[(1-a)/N]^q\Omega$,
so the power PGM has exactly the objective in
Eq.~\eqref{eq:sm:bk-upper-fidelity}.
Measuring in the syndrome basis and guessing the state $\kb{e_i}{e_i}$
for outcome $e_i$ with $i\ge1$ again gives $f^\opt\geq a$. Here
$\kappa=T=c_\zeta=1$, and the same
bound therefore proves a strict violation for a finite ensemble of pure
states.
\end{proof}

\section{Sectored orthogonal channels and maximum likelihood decoding}
\label{sec:sm:soc}
For a channel satisfying Eq.~\eqref{eq:soc}, the uniform reference $\zeta=\tau$
gives $G_q=d_AC^q$. We use the MLD in Eq.~\eqref{eq:mld-kraus} and
the common completion outside the populated sectors specified in
Sec.~\ref{sec:sm:basic}.

\begin{smtheorem}[Maximum likelihood limit]
\label{thm:sm:mld}
For $\ECal\in\SOC$ and $\zeta=\tau$, the channel fidelity $f^{[q]}$ is
nondecreasing as a function of $q>0$, with
\begin{equation}\label{eq:sm:soc-power}
 f^{[q]}=\sum_s\frac{\sum_\ell\mu_{s\ell}^{q+1}}
                         {\sum_\ell\mu_{s\ell}^{q}}
 \longrightarrow f^{\opt}=\sum_s \mu_{s,\max}
\end{equation}
as $q\to\infty$. The PPM decoder converges in diamond norm:
\begin{equation}\label{eq:sm:soc-mld-limit}
 \lim_{q\to\infty}\norm{\PCal_{\tau,\widehat\FCal^{[q]}}-\RCal^{\MLD}}_\diamond=0.
\end{equation}
\end{smtheorem}
\begin{proof}
The vectors $\dket{E_{s\ell}}/\sqrt{d_A}$ are orthonormal eigenvectors
of $J(\ECal)/d_A$, with eigenvalues $\mu_{s\ell}$ and output marginals
$\proj_s/d_A$. Substituting the spectral decomposition of $J(\ECal)/d_A$ into
Eq.~\eqref{eq:sm:normalization} gives the decoder on the populated sectors,
\begin{equation}\label{eq:sm:soc-decoder}
 \PCal_{\tau,\widehat\FCal^{[q]}}(\omega)=\sum_{s,\ell}
 \frac{\mu_{s\ell}^q}{\sum_m\mu_{sm}^q}E_{s\ell}^\dagger\omega E_{s\ell}.
\end{equation}
Trace orthogonality in Eq.~\eqref{eq:soc} gives the fidelity in
Eq.~\eqref{eq:sm:soc-power} and shows that the MLD attains
$\sum_s\mu_{s,\max}$. Conversely,
\begin{equation}\label{eq:sm:soc-dual}
 C\preceq I_A\otimes \Gamma_{\ECal_{\mathrm{SOC}}},\qquad
 \Gamma_{\ECal_{\mathrm{SOC}}}\coloneqq \frac1{d_A}\sum_s\mu_{s,\max}\proj_s,
\end{equation}
because the Choi matrix eigenvectors in sector $s$ lie in
$\HCal_A\otimes\supp(\proj_s)$. The matching dual trace proves optimality.
To prove monotonicity, fix a sector $s$ and define
$p_{s\ell}\coloneqq\mu_{s\ell}^q/\sum_m\mu_{sm}^q$.
Holding the channel weights $\mu_{s\ell}$ fixed, we differentiate with respect to $q$ to obtain
\begin{equation}\label{eq:sm:soc-monotonicity}
 \begin{aligned}
 \frac{\mathrm{d}}{\mathrm{d}q}\left(\frac{\sum_\ell\mu_{s\ell}^{q+1}}
 {\sum_m\mu_{sm}^q}\right)
 &=\sum_\ell\mu_{s\ell}\frac{\mathrm{d}p_{s\ell}}{\mathrm{d}q}\\
 &=\sum_\ell p_{s\ell}\mu_{s\ell}
 \left(\log\mu_{s\ell}-\sum_m p_{sm}\log\mu_{sm}\right)\\
 &=\frac12\sum_{\ell,m}p_{s\ell}p_{sm}
 (\mu_{s\ell}-\mu_{sm})(\log\mu_{s\ell}-\log\mu_{sm})\ge0.
 \end{aligned}
\end{equation}
The logarithm is increasing, so each product of differences in the last
sum is nonnegative. Summing over sectors proves that $f^{[q]}$ is
nondecreasing.
Finally, divide the numerator and denominator of each coefficient in
Eq.~\eqref{eq:sm:soc-decoder} by $\mu_{s,\max}^q$.
This proves convergence to the uniform mixture over the maximizing
classes. Because the coefficients converge within a fixed finite family
of Kraus operators, the convergence also holds in diamond norm, as stated in
Eq.~\eqref{eq:sm:soc-mld-limit}.
\end{proof}

\subsection{Finite Abelian stabilizer examples}
Let $G$ be a finite Abelian configuration group, let $\widehat G$ be its
character group, and take $\HCal_B=\mathbb C[G]$.
The Weyl operators $D_z$, indexed by $z\in G\times\widehat G$, combine
basis shifts with character phases and form an orthogonal unitary
operator basis~\cite{BermejoVegaVanDenNest2014,Albert2020}.
Choose a commuting subgroup $\SCal$ of Weyl operators, with scalar
phases fixed so that its common $+1$ eigenspace is nonzero. Let
$V:\HCal_A\to\HCal_B$ be an
isometric encoding whose range is this entire eigenspace. Its projector
is
\begin{equation}\label{eq:sm:abelian-code-projector}
 P\coloneqq VV^\dagger=\frac1{|\SCal|}\sum_{S\in\SCal}S,
 \qquad d_A=\frac{|G|}{|\SCal|}.
\end{equation}
For a probability distribution $\{p_z\}$, stochastic Weyl noise is
$\NCal(\omega)\coloneqq\sum_zp_zD_z\omega D_z^\dagger$.
The encoded channel has Kraus operators $\sqrt{p_z}D_zV$.
The commutation characters with $\SCal$ determine the syndrome.
Operators with distinct syndromes have orthogonal ranges, whereas those
with the same syndrome have the same range. Moreover,
\begin{equation}\label{eq:sm:abelian-overlap}
 \tr[(D_zV)^\dagger(D_wV)]=\tr(PD_z^\dagger D_w).
\end{equation}
Weyl trace orthogonality makes this expression vanish unless
$D_z^\dagger D_w$ is proportional to a stabilizer. In that case, the
restricted isometries agree up to a phase. Choose one representative
for each class of equivalent isometries and add their probabilities.
After we omit classes of zero probability, the resulting isometries
$E_{s\ell}$ and weights $\mu_{s\ell}$ satisfy
Eq.~\eqref{eq:soc}. Thus every channel of this form is an
SOC\@.

This inclusion relies on both stochastic Weyl noise and a stabilizer
subspace code. A coherent superposition of Weyl errors, or a general
code described only by group labels, need not have the same structure.
Finite-energy regularization of oscillator GKP codes need not preserve
exact sector orthogonality~\cite{GKP2001}.

\section{Numerical examples of large power recovery}
\label{sec:sm:figure-models}
\begin{figure}[t]
\centering
\includegraphics[width=0.98\textwidth]{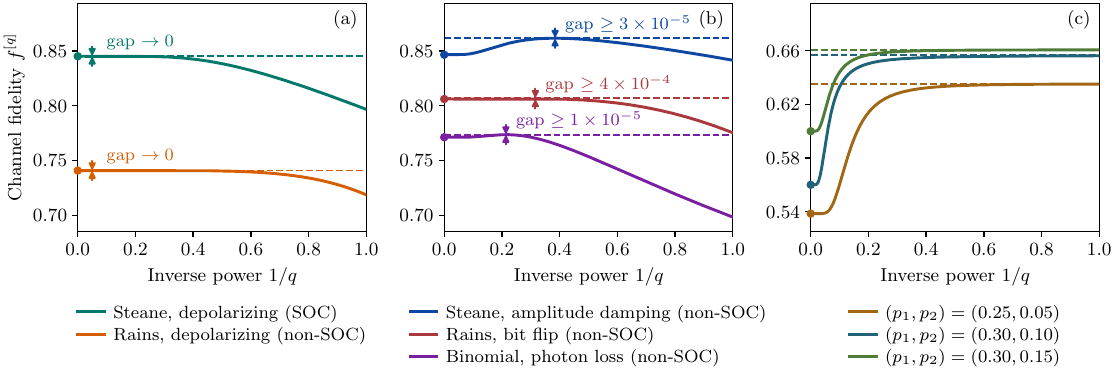}
\caption{Channel fidelity $f^{[q]}$ versus inverse power $1/q$.
Power increases from right to left, and the dots at $1/q=0$ mark the
limits. For each channel, a horizontal dashed line in the same color
marks the optimal recovery fidelity $f^\opt$.
(a) Under independent depolarizing noise, the Steane code
($p=0.12$, SOC) and the Rains code ($p=0.08$, non-SOC) approach their
respective optima. (b) The Steane code under amplitude damping
($\gamma_d=0.3$), the Rains code under independent bit flips ($p=0.1$),
and the binomial code under photon loss ($\gamma_{\mathrm{loss}}=2/5$)
retain positive limiting gaps. Arrows in (b) mark the
smallest numerical gaps, rounded down to one significant digit.
(c) For all three parameter pairs of the flagged qubit channel in
Eq.~\eqref{eq:sm:flagged-qubit}, fidelity decreases as $q$ increases from one.}
\label{fig:sm:recovery-models}
\end{figure}

Fig.~\ref{fig:sm:recovery-models} compares recovery for three quantum
codes and a qubit channel with a quantum flag. All examples use channel
fidelity at the maximally mixed input $\tau$. For the code examples,
$\ECal=\NCal\circ\CCal$ consists of the isometric encoding
$\CCal(\omega)=V\omega V^\dagger$ followed by the specified noise
$\NCal$.

\subsection{Asymptotically optimal recovery inside and outside SOC}
For the $[[7,1,3]]$ Steane code~\cite{Steane1996}, independent Pauli
depolarizing noise acts on each physical qubit as
\begin{equation}\label{eq:sm:steane-depolarizing}
 \DCal_p(\omega)=(1-p)\omega+
 \frac p3\sum_{\alpha=x,y,z}\sigma_\alpha\omega\sigma_\alpha.
\end{equation}
Grouping errors by syndrome and logical Pauli class gives an SOC for
every $p\in[0,1]$. By Theorem~\ref{thm:sm:mld}, the fidelity in
Eq.~\eqref{eq:sm:soc-power} is nondecreasing in $q$ and converges to
its optimal value.

The nonadditive $((5,6,2))_2$ Rains code~\cite{Rains1997} under the same
independent noise with $p=0.08$ gives the second example in
Fig.~\ref{fig:sm:recovery-models}(a). This channel is outside SOC because
its populated output dimension $32$ is not divisible by the logical
dimension $6$. Nevertheless, its limiting fidelity is approximately
$0.740815$ and agrees with the numerical optimum to within
$4\times10^{-79}$ under 160-digit arithmetic, as verified by primal and dual bounds.
Thus asymptotically optimal recovery also occurs
outside the sufficient SOC structure.

\subsection{Persistent gaps outside SOC}
The first example in Fig.~\ref{fig:sm:recovery-models}(b) uses the same
Steane code with independent amplitude damping noise. Its single-qubit
Kraus operators are
$A_0=\kb{0}{0}+\sqrt{1-\gamma_d}\kb{1}{1}$ and
$A_1=\sqrt{\gamma_d}\kb{0}{1}$, with $\gamma_d=0.3$.
The second example uses the Rains code from panel (a), now with
independent bit flips on each physical qubit. The single-qubit Pauli
probabilities are $(p_I,p_X,p_Y,p_Z)=(1-p,p,0,0)$ with $p=0.1$.
The bit flips populate all $32$ output dimensions, so the same dimension
argument as in panel (a) rules out SOC. The third example uses the
simplest single-mode binomial
code~\cite{Michael2016binomial},
\begin{equation}\label{eq:sm:binomial-code}
 \ket{0_L}=\frac{\ket0+\ket4}{\sqrt2},\qquad
 \ket{1_L}=\ket2,
\end{equation}
under photon loss with loss probability $\gamma_{\mathrm{loss}}=2/5$.
Only Fock levels $0$ through $4$ are populated, so the five-dimensional
output representation is exact.

All three channels in panel (b) attain their highest observed fidelity
at a finite power. For the Steane, Rains, and binomial codes, the
numerical maxima occur at $q\simeq2.603$, $3.167$, and $4.677$,
respectively. Their gaps to the optimal recovery are approximately
$3.70\times10^{-5}$, $4.22\times10^{-4}$, and $1.08\times10^{-5}$.
As $q\to\infty$, the respective gaps grow to $1.51\times10^{-2}$,
$6.08\times10^{-4}$, and $2.40\times10^{-3}$.

The curves use $401$ uniformly spaced values of $1/q$, supplemented
by the observed interior maxima in panel (b). For the Steane code under
amplitude damping, the endpoint is a numerical limit checked at
$q=100,400,1000,2000$. The other endpoints follow from the limiting
decoders. Independent primal and dual bounds bracket the optimal
fidelities in panel (b) to within $1.5\times10^{-10}$ for amplitude
damping, $1.0\times10^{-12}$ for bit flips, and $8.3\times10^{-13}$
for photon loss.

\subsection{Decreasing fidelity for a flagged qubit channel}
The channel in Fig.~\ref{fig:sm:recovery-models}(c) applies the identity
to the transmitted qubit with probability $p_1$ and applies each of the
qubit Pauli matrices $\sigma_x$ and $\sigma_z$ with probability
$(1-p_1)/2$. The state of the output flag qubit depends on the operation.
With the flag listed first, the Kraus operators $K_i\colon\mathbb C^2\to\mathbb C^4$ are
\begin{equation}\label{eq:sm:flagged-qubit}
 \begin{gathered}
 K_0=\sqrt{p_1}\ket0\otimes I_2,\qquad
 K_+=\sqrt{\frac{1-p_1}{2}}\ket{h_+}\otimes\sigma_x,\qquad
 K_-=\sqrt{\frac{1-p_1}{2}}\ket{h_-}\otimes\sigma_z,\\
 \ket{h_\pm}=\sqrt{p_2}\ket0\pm\sqrt{1-p_2}\ket1.
 \end{gathered}
\end{equation}
Here $p_2$ is the squared overlap of either error flag with $\ket0$.
For $(p_1,p_2)=(0.25,0.05)$, $(0.30,0.10)$, and $(0.30,0.15)$, the
gaps $f^\opt-f^{[1]}$ are approximately $1.38\times10^{-5}$,
$3.98\times10^{-4}$, and $4.42\times10^{-6}$, respectively.
For each pair, $f^{[q]}$ decreases strictly as $q$ increases from one.

\end{document}